\documentclass{article}
\usepackage{graphicx}
\usepackage[modulo, displaymath, mathlines]{lineno}
\usepackage{epsfig, mathtools, caption}
\usepackage{multicol, fullpage, natbib}
\usepackage[section]{placeins}
\usepackage{threeparttable}

\title{Dilatancy stabilises shear failure in rock}
\author{Franciscus M. Aben, 
Nicolas Brantut\\
  Department of Earth Sciences\\
University College London, UK}
\date{}

\begin{document}
\maketitle

%\linenumbers
\begin{abstract}
Failure and fault slip in crystalline rocks is associated with dilation. When pore fluids are present and drainage is insufficient, dilation leads to pore pressure drops, which in turn lead to strengthening of the material. We conducted laboratory rock fracture experiments with direct in-situ fluid pressure measurements which demonstrate that dynamic rupture propagation and fault slip can be stabilised (i.e., become quasi-static) by such a dilatancy strengthening effect. We also observe that, for the same effective pressures but lower pore fluid pressures, the stabilisation process may be arrested when the pore fluid pressure approaches zero and vaporises, resulting in dynamic shear failure. In case of a stable rupture, we witness continued prolonged slip after the main failure event that is the result of pore pressure recharge of the fault zone. All our observations are quantitatively explained by a spring-slider model combining slip-weakening behaviour, slip-induced dilation, and pore fluid diffusion. Using our data in an inverse problem, we estimate the key parameters controlling rupture stabilisation, fault dilation rate and fault zone storage. These estimates are used to make predictions for the pore pressure drop associated with faulting, and where in the crust we may expect dilatancy stabilisation or vaporisation during earthquakes. For intact rock and well consolidated faults, we expect strong dilatancy strengthening between 4 and 6 km depth regardless of ambient pore pressure, and at greater depths when the ambient pore pressure approaches lithostatic pressure. In the uppermost part of the crust ($<4$ km), we predict vaporisation of pore fluids that limits dilatancy strengthening. The depth estimates where dilatant stabilisation is most likely coincide with geothermal energy reservoirs in crystalline rock (typically between 2 and 5 km depth) and in regions  where slow slip events are observed (pore pressure that approaches lithostatic pressure). 
\end{abstract}

\section{Introduction}
Pore fluids are ubiquitous throughout the earth's crust, and may govern or affect fault dynamics in settings ranging from subduction zones, crustal scale strike-slip faults to geothermal reservoirs. Pore fluid pressure is controlled by changes in pore volume, which can be induced by deformation and fault slip. One key pore volume change phenomenon is dilatancy, which is the increase of the pore volume in rocks by small-scale brittle deformation (e.g., microfractures, opening of pore space between grains), and is expected to be significant in rocks with a low porosity and well consolidated fault gouges. The added pore volume acts as a pore pressure sink when the fault zone is partially or entirely undrained -- a scenario likely to occur during faulting, when new pore volume is created at a rate larger than the rate of fluid recharge from outside the fault zone. A drop in fault zone pore pressure increases the normal stress on the fault, thereby increasing the shear resistance. Dilatancy strengthening is a transient phenomenon, as the extraneous shear resistance vanishes when pore pressure recovers to its initial value over time. Theoretical work over the past four decades \citep[e.g.,][]{Rice1979b, Rudnicki1988, Segall1995, Segall2010, Ciardo2019} has shown that dilatancy strengthening can potentially stabilise an otherwise (i.e., under dry or drained conditions) unstable fault by suppressing or delaying acceleration of slip, which has primary consequences for earthquake nucleation and rupture dynamics, and increases our ability to observe potential earthquake precursory signals \citep[e.g.][]{Nur1972, Scholz1973, Bouchon2013, Shreedharan2021}. Despite the central role played by dilatancy in the mechanics of faulting, theoretical predictions have remained largely untested and only indirect, qualitative experimental evidence of rupture and slip stabilisation in granite \citep{Martin1980} and fault gouge \citep{Lockner1994, Xing2019, Proctor2020} have been brought forward. 

Two main obstacles remain to assess quantitatively the impact of dilatancy on earthquake nucleation and slip. Firstly, direct experimental evidence of dilatancy-induced rupture stabilisation is still lacking, and it remains unclear whether the common model assumptions are a correct representation of the reality of hydro-mechanical coupling during rock failure and fault slip. Secondly, quantitative measurements of key model parameters under in-situ conditions are still not available. Indeed, most theoretical studies emphasise the lack of experimental measurements on the effectiveness of a shear fault zone to act as a pore pressure sink, which severally impedes application of their theoretical understanding to make meaningful predictions of dilatancy stabilisation. Two parameters are required to describe a fault's effectiveness as a pore pressure sink: The dilation rate, i.e., how porosity increases with fault slip and other evolving quantities, and the so-called storage capacity of the fault zone material, which is a compressibility that relates the pore volume change to the change in pore pressure.

Dilation rates have been measured during slip on natural and simulated fault gouges \citep[e.g.,][]{Marone1990,Lockner1994,Samuelson2009}, after shear failure of intact rock \citep[e.g.,][]{Rummel1978}, and during reactivation by pore fluid injection of a natural fault zone \citep{Guglielmi2015}. By contrast, fault zone storage is very difficult to measure. Fault zone storage describes how changes in pore pressure are buffered by compression/extension of the pore fluid and the pore space. It can be estimated from the storage capacity $S_\mathrm{f}$ of the dilated fault zone material (given by the compressibilities of the pore space and pore fluid times the porosity \citep[e.g.,][]{Jaeger2007}) multiplied by the width of the fault zone disturbed by dilatancy. Measuring $S_\mathrm{f}$ and the disturbed fault zone width is challenging: For fault gouges, storage capacities have been measured only in an undisturbed state \citep[e.g.,][]{Wibberley2002}, and the width of the disturbed gouge layer in natural fault zones may vary over some orders of magnitude even during a slip event \citep[see discussion in][]{Rice2006}. Measurements of the local pore pressure in the fault zone \emph{during} fault slip are required to estimate the effective fault zone storage. Previous experimental studies that report on dilatancy stabilisation in intact rock \citep{Martin1980} and stick-slip events in consolidated gouge \citep{Lockner1994, Xing2019} lacked the on-fault pore pressure measurements to calculate the fault zone storage. Recent advances in laboratory instrumentation \citep[e.g.,][]{Brantut2021} allow for direct measurements of the pore pressure drop caused by dilatancy during shear failure of intact rock \citep{Brantut2020} and during slip on a saw-cut granite surface and in a gouge layer \citep{Proctor2020}. Such new techniques open the way for systematic studies of dilatancy strengthening phenomena, including direct tests of model predictions and quantitative parameter estimation.

Here, we leverage our newly developed in-situ pore pressure measurement technique \citep{Brantut2020,Brantut2021} to show direct evidence of rupture stabilisation by dilatancy strengthening, estimate the key parameters necessary to quantify dilatancy strengthening in faults, and to predict at which depths and pore pressure conditions dilatancy strengthening can indeed cause stabilisation of shear failure. 

\section{Results}
The experimental setup used in this study allows us to measure local pore pressure changes at several locations on the surface of a cylindrical rock sample during deformation in a conventional triaxial loading apparatus. The local pore pressure changes were measured with in-house developed miniature pore pressure sensors \citep{Brantut2021}. We used Westerly granite cylindrical samples with two opposite facing notches cut at a $30^\circ$ angle with respect to the sample axis, in order to have prior knowledge of the prospective failure plane during axial loading (Figure \ref{fig:stress-strain}a). The samples were thermally cracked at $600^\circ$C, and intact hydraulic properties of the samples were characterised prior to deformation (see Methods and Figure \ref{fig:permstorativity}). The samples were deformed at nominal effective pressures of 40 MPa and 80 MPa, with a varying combination of confining pressure $P_\mathrm{c}$ and imposed pore pressure $p_0$ (Table \ref{tab:1}). The imposed pore pressure during deformation was kept constant at both sample ends. Axial load was increased by imposing a constant overall deformation rate of $10^{-6}$~s$^{-1}$. The samples were loaded until shear failure, after which the piston displacement was arrested until complete reequilibration of the pore pressure, so that the total pore volume change due to failure in the sample could be measured. Deformation was subsequently resumed, and slip continued on the newly formed fault. Periods of stable sliding or stick-slip events occurred. After each slip increment, we paused deformation to remeasure pore volume changes. More details are provided in the Methods section.

\begin{table}
  \centering
  \begin{threeparttable}[b]
    \caption{Summary of shear failure experiments. Nominal strain rate was $10^{-6}$ s$^{-1}$ for all experiments.}
    \label{tab:1}
    \begin{tabular}{ccccccccccc}
      \hline
      & Effective & Confining & Pore &  & Min. pore & Peak slip & Dilation & Storage & Slip\\
                 &  pressure &  pressure &  pressure & Failure &  pressure &  rate &  rate & capacity & weakening\\ 
      Rock & $P_\mathrm{eff}$ & $P_\mathrm{c}$ & $p_0$ & mode & $p_\mathrm{min}$ & $v_\mathrm{max}$ & $dw/d\delta$ & $S_\mathrm{f}w$ & $\delta_\mathrm{c}$\\
           & (MPa) & (MPa) & (MPa) &  & (MPa) & (mm s$^{-1}$) & (-) & (mm GPa$^{-1}$) & (mm) \\
      \hline
      WG6 & $40$ & $40$ & $0$ (dry) & stable\tnote{$\dagger$} & - & - & - & - & - \\
      WG3\tnote{a} & $40$ & $60$ & $20$ & dynamic & $0$\tnote{b} & $>10$\tnote{c} & $0.094$ & - & - \\
      WG2\tnote{a} & $40$ & $70$ & $30$ & dynamic & $0$\tnote{b} & $>10$\tnote{c} & - & - & - \\
      WG5  & $40$ & $100$ & $60$ & stable & $26$ & $0.74$ & $0.146$ & $3.84$ & $1.68$ \\
      WG12 & $40$ & $110$ & $70$ & stable & $42$ & $0.27$ & $0.171$ & $5.93$ & $2.01$ \\
      WG4  & $40$ & $120$ & $80$ & stable & $37$ & $1.23$ & $0.082$ & $1.65$ & $1.84$ \\
      WG10 & $80$ & $120$ & $40$ & dynamic & $0$\tnote{b} & $>10$\tnote{c} & $0.054$ & - & - \\
      WG7  & $80$ & $160$ & $80$ & dynamic & $<14$\tnote{d} & $>10$\tnote{c} & $0.069$ & - & - \\
      WG8  & $80$ & $160$ & $80$ & stable & $16$ & $0.65$ & $0.077$ & $1.28$ & $1.87$ \\
      \hline 
    \end{tabular}
    \begin{tablenotes}
    \item[$\dagger$] Stable rupture was achieved by controlling acoustic emission rate as in \citet{Aben2019}.
    \item[a] From \citet{Brantut2020}.
    \item[b] Pore water at partially vaporised or degassed.
    \item[c] $10$~mm~s$^{-1}$ is the maximum detectable value in our system.
    \item[d] Pore pressure transducers out of range.
    \end{tablenotes}
  \end{threeparttable}
\end{table}

\subsection{Shear failure of intact rock}
\begin{figure}
\centering
\includegraphics{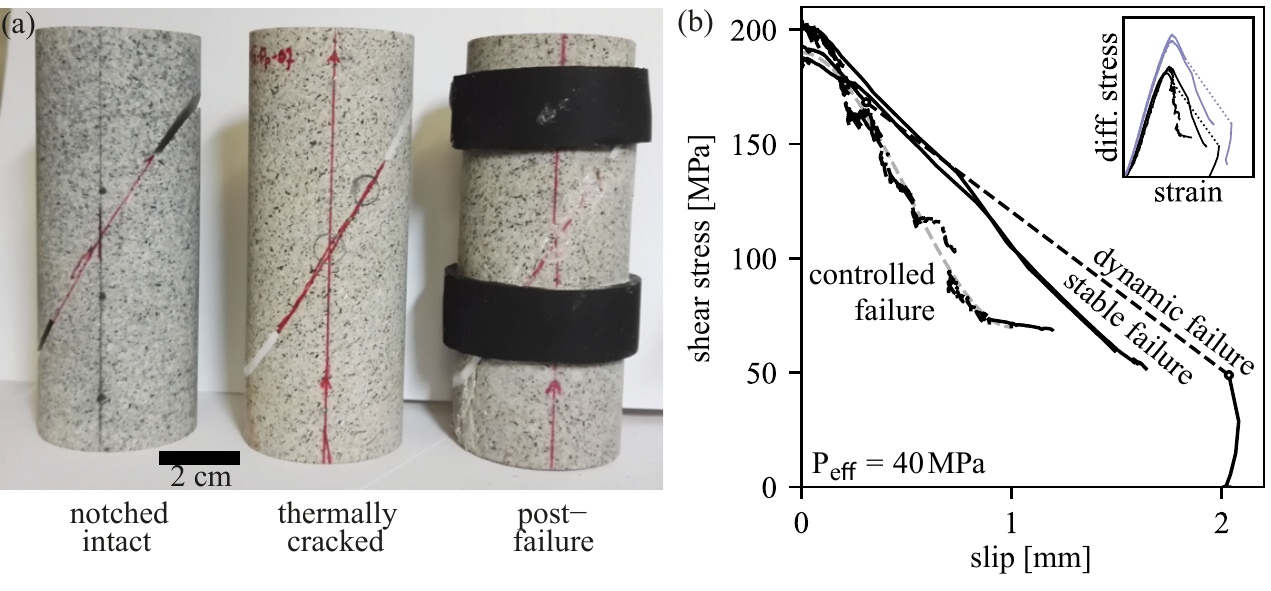}
\caption{\textbf{(a)}: Notched Westerly granite samples that are intact (left), thermally cracked (center) and thermally cracked and failed under triaxial loading conditions (right). \textbf{(b)}: Shear stress versus slip during stable and dynamic shear failure experiments performed at 40~MPa effective pressure. Dynamic failure is indicated by the dashed portion of the curve. Dry quasi-static failure at 40~MPa effective pressure was achieved by acoustic emission rate controlled loading feedback, and provides a direct estimate of the minimum breakdown work necessary to reach the residual frictional strength of the fault. Gray dashed line: Cohesion-weakening function $f(\delta)$ used in spring-slider simulations. Inset: Some stress versus strain curves for stable, dynamic, and controlled shear failure at 40~MPa (black curves) and at 80~MPa (purple curves) effective pressure. }
\label{fig:stress-strain}
\end{figure}

Under the pressure and temperature conditions of our tests, thermally cracked Westerly granite is brittle and when conducted under dry conditions deformation is always characterised by dynamic failure along a shear fault. Here, in our water saturated experiments, two types of behaviour were observed depending on the initial pore pressure $p_0$ (Table \ref{tab:1}). At $P_\mathrm{eff}=40$~MPa, tests conducted at $p_0=20$ and $30$~MPa underwent dynamic shear failure. At $P_\mathrm{eff}=80$~MPa, the test conducted at $p_0=40$~MPa and one test where $p_0=80$~MPa also failed dynamically. Dynamic failures are characterised by the well-known `bang'. We first observed a slow onset of the reduction in stress beyond the peak fracture strength that accelerated strongly. The majority of the stress reduction during dynamic failure was accommodated within less than a second (Figure \ref{fig:stress-strain}b, dashed segment). No clear deceleration in the stress reduction rate was observed. 

By contrast, at initially high pore pressure ($p_0=60$ to $80$~MPa and $P_\mathrm{eff}=40$~MPa, and one test at $p_0=80$~MPa and $P_\mathrm{eff}=80$~MPa), sample failure occurred silently, which indicates that failure remained quasistatic. The entire stress drop took several minutes, and the largest reduction in stress occurred in a timespan of the order of tens of seconds. The peak slip rate was typically less than $1$~mm~s$^{-1}$. We term these ruptures ``stable''. 

A clear shear fault formed between the notches for most experiments (e.g., Figure \ref{fig:stress-strain}a), except for two samples (deformed at $P_\mathrm{c}$ = 100~MPa, $p_0$ = 60~MPa, and  $P_\mathrm{c}$ = 110~MPa, $p_0$ = 70~MPa) where the fault plane deflected from its prospective path towards the sample's end.

\begin{figure}
\centering
\includegraphics{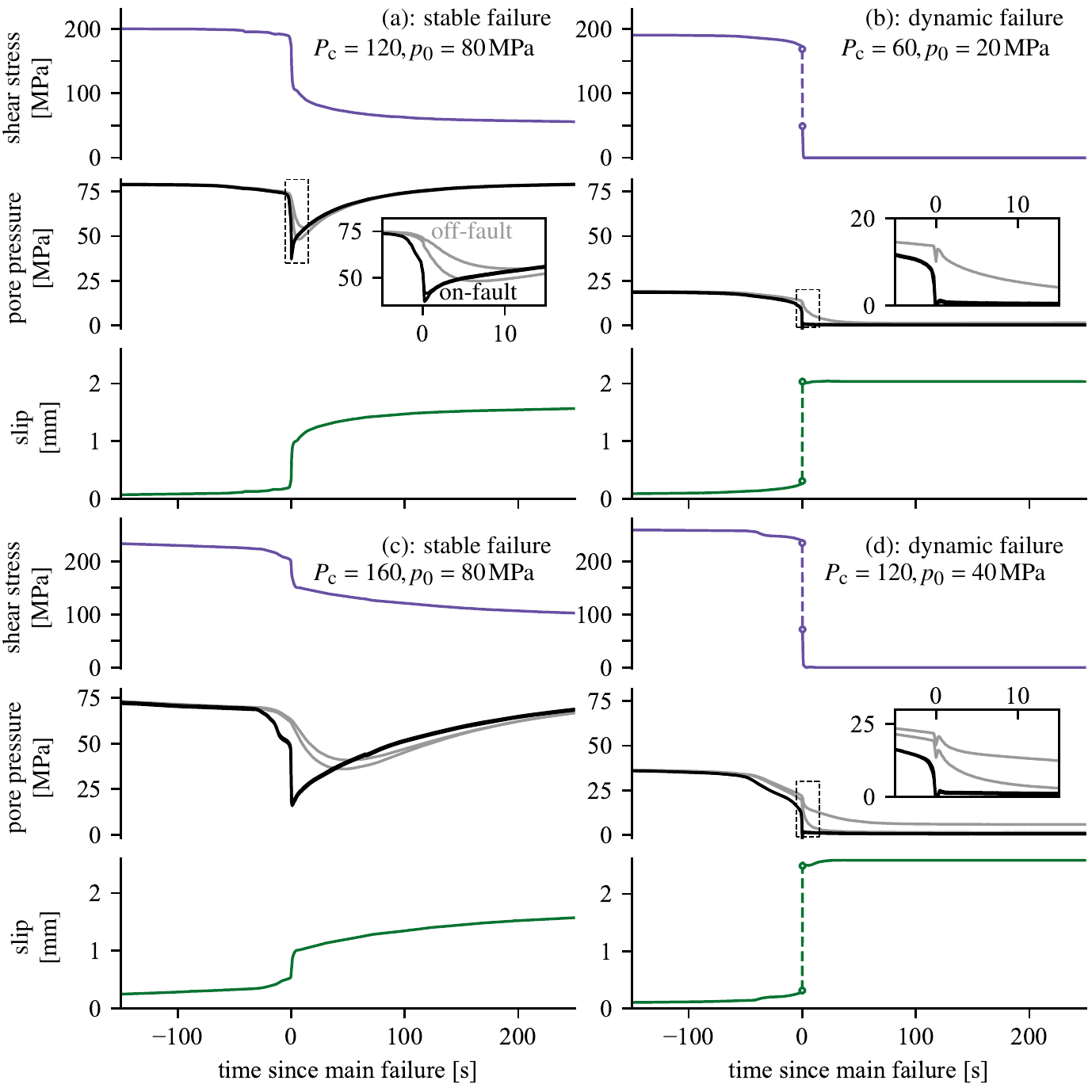}
\caption{Records of shear stress (top), pore pressure (middle), and fault slip (bottom) for stable failure at \textbf{(a)}: $P_\mathrm{eff}=40$~MPa (sample WG4) and at \textbf{(b)}: $P_\mathrm{eff}=80$~MPa (sample WG8), and for dynamic failure at \textbf{(c)}: $P_\mathrm{eff} = 40$~MPa (sample WG3) and at \textbf{(d)}: $P_\mathrm{eff}=80$~MPa (sample WG10). Interval for dynamic interval shown as a dashed curve. Middle panel: On-fault pore pressure records shown as black curves, off-fault pore pressure records shown as gray curves. Insets show pore pressures around the main failure event. Data from other samples are shown in Figure S2. }
\label{fig:porepressure}
\end{figure}

Pore pressure was recorded on the fault by two local pressure sensors, and recorded outside of the failure zone by two additional sensors (Figures \ref{fig:A},  \ref{fig:porepressure}). In all samples, the pore pressure remained uniform in the sample until the peak differential stress, at about one MPa lower than the imposed pore pressure (Figure \ref{fig:porepressure}). Beyond the peak stress, pore pressure decreased homogeneously throughout the sample at first, and then dropped more rapidly on-fault than off-fault. In both the dynamic and stable cases, the stress drop during failure was associated with a strong pore pressure drop on the fault, of the order of several tens of MPa. The off-fault pore pressure also decreased but in a more gradual manner, and with a delay.

The pore pressure evolution during and after failure was different between dynamic and stable ruptures. In all samples that failed dynamically, the on-fault pore pressure dropped to zero concurrently with the main shear stress drop, and remained constant for the subsequent 5 to 10 minutes, while deformation was stopped. The off-fault pore pressure did not drop to zero but also remained constant at very small values (a few MPa) for the same time period. The shear stress dropped immediately towards zero, overshooting the residual frictional strength of the rock (Figure \ref{fig:stress-strain}b) -- a typical observation for dynamic laboratory failures. The lack of an immediate pore pressure recovery in the fault zone can be interpreted as an indication for local vaporisation or degassing \citep{Brantut2020}.

By contrast, during stable failure, the on-fault pore pressure dropped to a minimum significantly above zero (typically tens of MPa) as slip rate reached it maximum and shear stress experienced a relatively rapid first drop. After this first stress drop, we stopped deformation by locking the axial piston in place. The pore pressure on the fault immediately increased, while shear stress decreased further and more fault slip was accumulated. The on-fault pore pressure then asymptotically recovered to the imposed $p_0$ at the boundary of the sample. The off-fault pore pressure experienced a gradual decrease during failure, followed by a recovery. The pore pressure recovery was slower off than on the fault, to the extent that the off-fault pore pressure was transiently lower than that measured on the fault, indicating more rapid recharge inside the fault zone than outside it.

\begin{figure}
\centering
\includegraphics{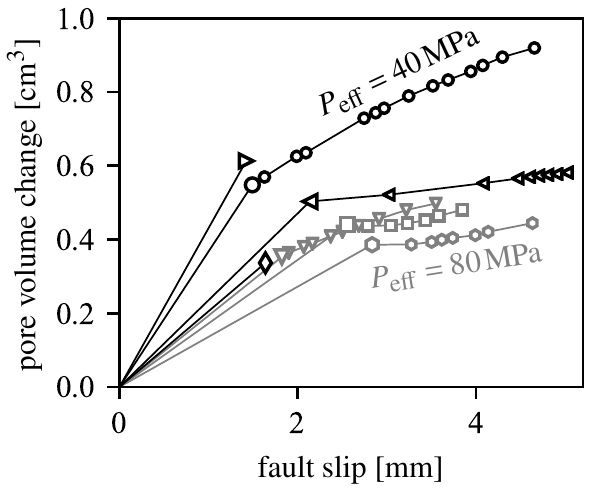}
\caption{Cumulative pore volume change versus fault slip for shear failure (large markers) and subsequent episodes of fault slip (small markers) in samples deformed at an initial effective pressure of 40~MPa (black curves) and 80~MPa (gray curves). Left-pointing triangle = WG3, diamond = WG4, circle = WG5, square = WG7, inverted triangle = WG8, hexagon = WG10, right-pointing triangle = WG12. }
\label{fig:porevolumechange}
\end{figure}

The total pore volume in the samples, as measured by the difference in intensifier volume before rupture and after complete pressure reequilibration, increased strongly during the failure of the intact rock (Figure \ref{fig:porevolumechange}). On average, additional pore volume created during failure at 80~MPa initial effective pressure is lower than at 40~MPa effective pressure. We express dilation as an increase in fault width $w$ as a function of fault slip $\delta$, which is computed from the measured pore volume change (see Method section). We find dilation rates $dw/d\delta$ between $0.055$ and $0.171$ for shear failure from an intact state (Table \ref{tab:1}). The two highest dilation rates were observed for samples where the fault curved away from one of the notches instead of following the intended trajectory, resulting in a more wavy fault zone (samples WG5 and WG12). For the twin experiments performed at the same imposed experimental conditions ($P_\mathrm{c}$ = 160~MPa, $p_0$ = 80~MPa), we observe a marginally higher dilation rate for the stable failure (sample WG8) relative to the dynamic failure (sample WG7).

\begin{figure}
\centering
\includegraphics{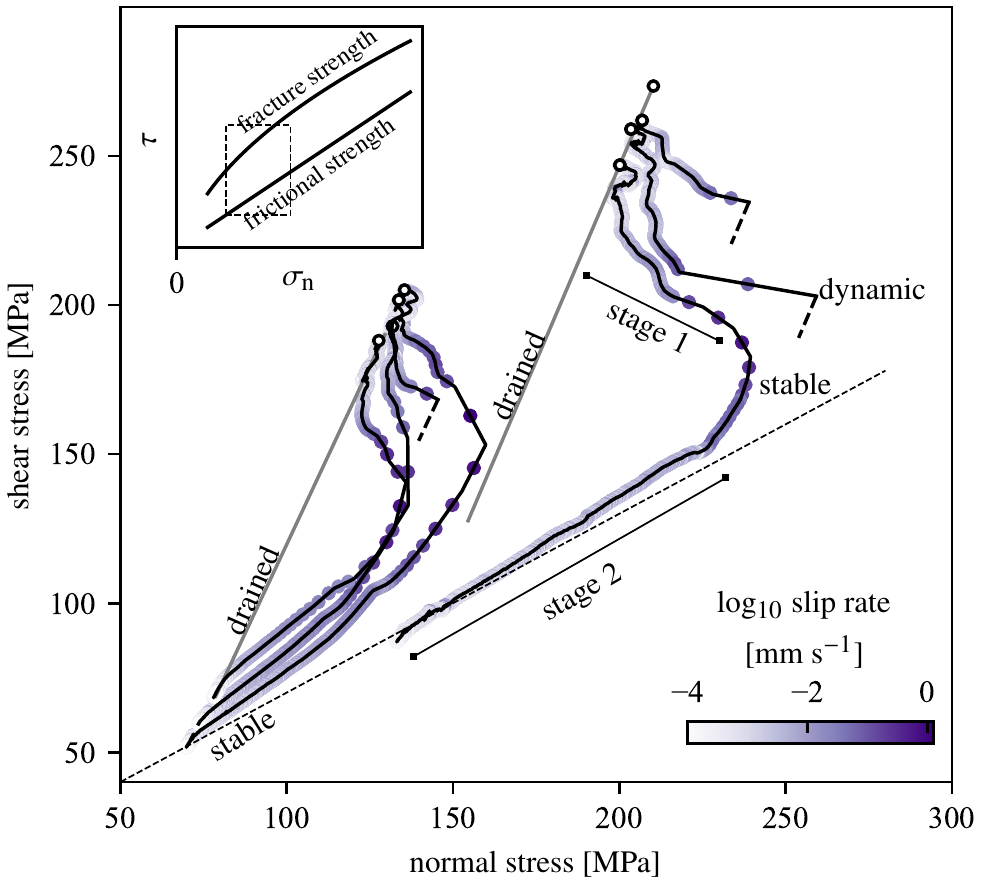} 
\caption{\textbf{(a)}: Shear failure stress paths from the fracture strength (peak shear stress) down to the residual frictional strength for dry or drained shear failure (gray curves), and for partially drained stable and dynamic shear failure (black curves, unstable experiments shown up to the onset of dynamic failure). Fault normal stress is corrected for the fault zone pore pressure measured during failure. Frictional strength for Westerly granite from \cite{Byerlee1967} (dashed line). The inset shows the overall trend of fracture and frictional strength \citep{Byerlee1967} and highlights the approximate range of our experiments in the main panel. }
\label{fig:Byerleeplot1}
\end{figure}

The large drop in pore pressure localised in the fault zone and the increase in pore volume provide direct evidence for strong fault zone dilatancy during shear failure. We can use the pore pressure recorded in the fault zone during shear failure to decouple mechanical weakening and pore pressure effects. We compute the effective normal stress on the fault, obtained as $\sigma_\mathrm{n} - p$, where $p$ is the measured on-fault pore pressure, and analyse the stress paths in a Mohr diagram (Figure \ref{fig:Byerleeplot1}). The unloading path for shear failure of Westerly granite under drained or dry conditions is shown for reference (Figure \ref{fig:Byerleeplot1}, gray curve). The unloading paths for both stable and dynamic shear failure under partially drained conditions deviate from the drained case. Considering stable failure first, we identify two consecutive stages in the unloading path: An increase in effective normal stress, directing the unloading path towards the frictional strength envelope for Westerly granite \citep{Byerlee1967,Lockner1998} (Stage 1, Figure \ref{fig:Byerleeplot1}), followed by a decrease in normal stress so that the unloading path closely follows the frictional strength envelope (stage 2). The fault slip rate increases during stage 1, and decreases during stage 2. Fault slip accumulated during this part of the shear stress drop increases nearly linearly with the recovery of the fault zone pore pressure towards the imposed pore pressure (Figure \ref{fig:recharge}a) and both the rate of pore pressure recharge and slip decrease with time (Figure \ref{fig:recharge}b, c). The onset of dynamic shear failure under partially drained conditions occurs during stage 1, where the dynamic unloading path runs parallel to the drained unloading path but at higher normal stress (dashed lines, Figure \ref{fig:Byerleeplot1}). 

\subsection{Fault slip by stable sliding and stick-slip events}
\begin{figure}
\centering
\includegraphics{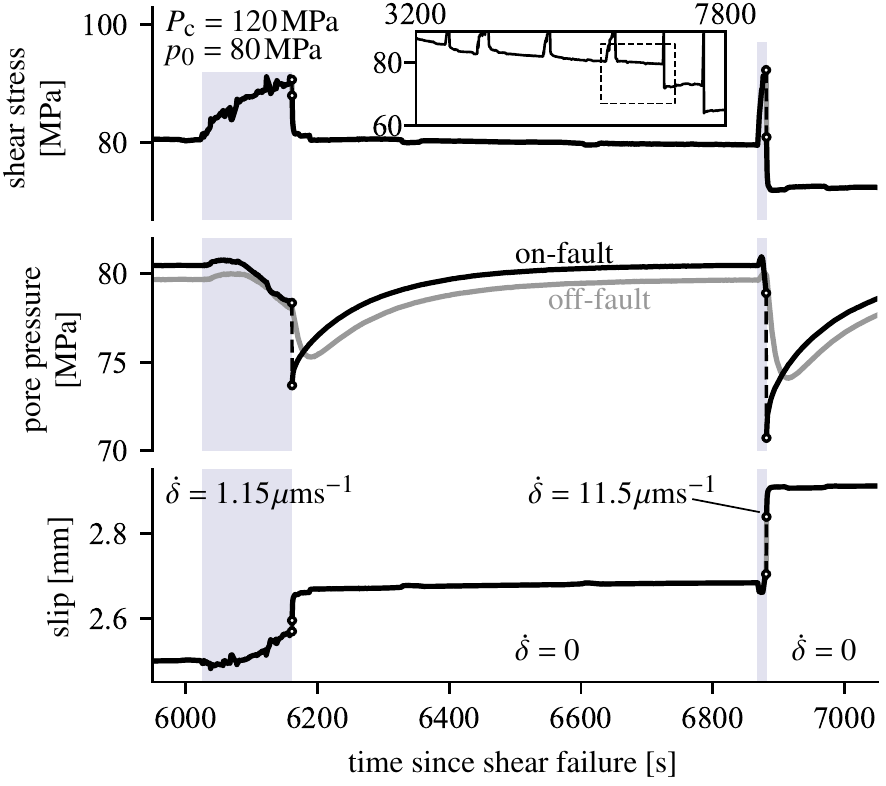} 
\caption{Records of the shear stress (top), pore pressure (middle), and fault slip (bottom) during progressive sliding on a fresh fault in granite. The records contains stick-slip events at imposed slip rates of 1.15 $\mu$m s$^{-1}$ and 11.5 $\mu$m s$^{-1}$(dynamic intervals shown by dashed curve). On-fault pore pressure records shown as black curves, off-fault pore pressure records shown as gray curves. Inset shows shear stress for the entire series of stick-slip events recorded for this sample (sample WG8). }
\label{fig:stickslip}
\end{figure}

\begin{figure}
\centering
\includegraphics{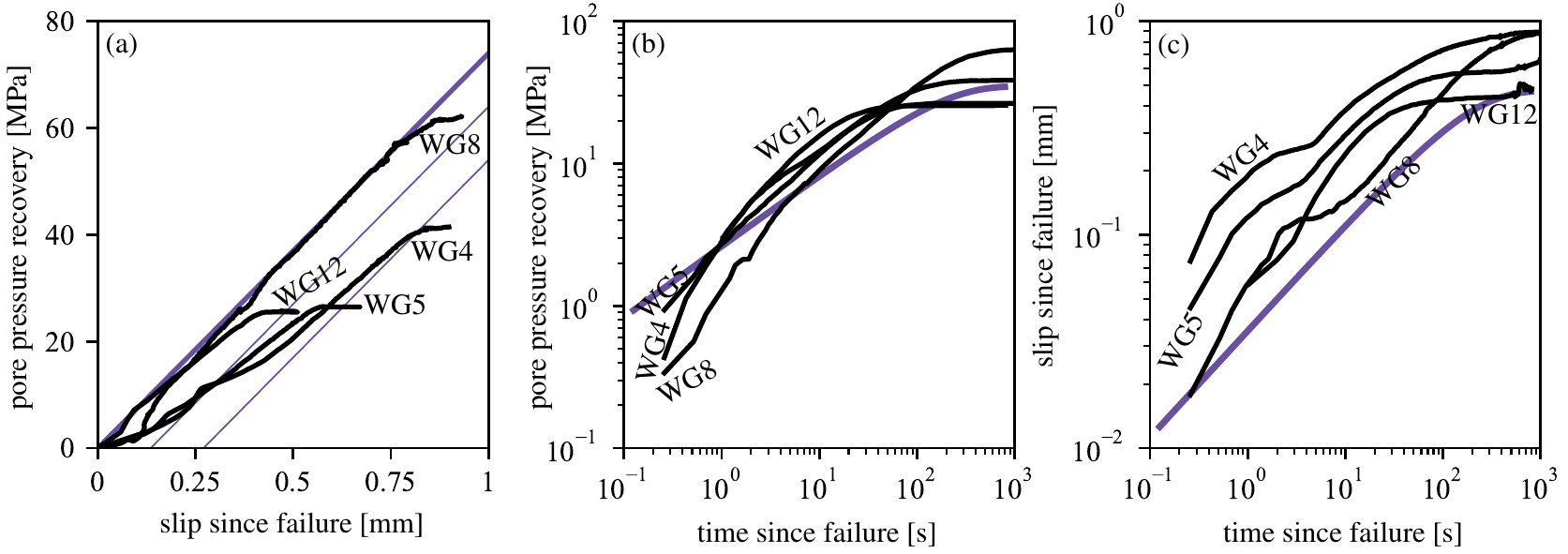} 
\caption{\textbf{(a)}: Pore pressure recovery $p$ in the fault zone after shear failure as a function of slip $\delta$, with $\delta = 0$ at $p = p_\mathrm{min}$. $p_\mathrm{min}$ is the lowest value for pore pressure reached during the failure process. Purple lines gives the relation between pore pressure and slip based on the unloading stiffness of the triaxial apparatus (equation \ref{eq:pslip}). \textbf{(b)}, \textbf{(c)}: Pore pressure recharge and fault slip versus time, shown from the lowest pore pressure reached during failure. Diffusion driven pore pressure recharge and slip are expected to scale with time as equation \eqref{eq:trech} (shown by purple curves).}
\label{fig:recharge}
\end{figure}

After shear failure, we imposed between 8 and 10 intervals of slip on the freshly formed fault at a slip rate of 1.15 $\mu$m s$^{-1}$ or 11.5 $\mu$m s$^{-1}$. We arrested the slip rate between each interval to allow pore pressure to reequilibrate and measure total pore volume change. For each slip interval, fault slip was accommodated either by stable sliding, by stick-slip, or by a combination of the two. We measured 6 to 8 stick-slip events per sample. The pore pressure initially increased at the onset of each interval, concurrently with an increase in shear stress (Figure \ref{fig:stickslip}). This was followed by a decrease in pore pressure, whilst shear stress approached either a new steady state value and fault slip remained stable, or the shear stress approached a peak value before dropping dynamically during a stick-slip event. During the stick-slip events, the pore pressure dropped by 1 to 10~MPa in the fault zone (Figure \ref{fig:stickslip}). After a stick-slip event, we observed a similar behaviour of pore pressure, fault slip, and shear stress to that observed after stable shear failure: Pore pressure recovered towards its imposed value, while shear stress decreases further and the fault continued to accumulate slip. The pore volume in the samples continued to increase during prolonged slip along the fault at a lesser rate than during shear failure (Figure \ref{fig:porevolumechange}). Dilation rates computed for sliding along the freshly created fault vary between $0.011$ and $0.046$, and do not vary between episodes of stable sliding or stick-slip. 

The initial increase of fault zone pore pressure suggests some compaction of the fault zone material from axial loading before the onset of dilation. The rise in shear strength and drop in pore pressure are indicative of dilatancy strengthening of the fault, but the strengthening effect is not sufficient to suppress unstable behaviour, as illustrated by the stick-slip events. These observations agree with recent observations of shear stress and local pore pressure in a simulated gouge layer and a bare granite saw cut experiment subject to imposed slip by \citet{Proctor2020}. 

\FloatBarrier

\section{Discussion}
Our experiments demonstrate that dilatancy strengthening leads to stabilisation of the shear failure process of intact rock by increasing the effective normal stress on the fault, whereas shear failure of Westerly granite under dry or drained conditions is typically unstable. Our data also shows a transition from stable to dynamic shear failure at the same effective pressure, which we propose results from insufficient initial pore pressure leading to pore fluid vaporisation or degassing, which in turns caps the efficiency of dilatancy strengthening and leads to dynamic failure. These observations are similar to those of \citet{Martin1980} on stable failure of granite at higher pore pressure, but we have unique measurements of the pore pressure evolution during stable and dynamic shear failure. These measurement indicate that stable shear failure consists of two consecutive stages: A stage dominated by a strong increase in fault strength by the dilatancy-induced pore pressure drop, followed by a stage of continued sliding during pore pressure recharge. As the latter stage follows the frictional strength envelope of Westerly granite, we conclude that the fault itself must have formed during the first stage. Stable failure thus effectively separates the loss of cohesion of intact rock from frictional sliding on a fresh fault. These findings on dilatancy strengthening during faulting have implications for earthquake nucleation and rupture dynamics that we shall discuss in this section. 

Ultimately, stability of rock failure in the laboratory is determined by the rock and machine stiffness and by the weakening of the fault zone material, the latter being impacted by pore pressure. In the following, we present a simple fault model which includes those key ingredients, and use it to fit our data and retrieve quantitative estimates of hydro-mechanical parameters, allowing us to make predictions for stable shear failure in the crust. 

\subsection{Coupled model and estimation of effective fault zone storage}
Prediction of dilatancy strengthening requires knowledge on the pore pressure drop in the part of the fault zone disturbed by dilatancy. For undrained conditions, where fluids within the fault zone are hydrologically isolated from their surroundings, the undrained pore pressure drop $\Delta p_\mathrm{undrained}$ is simply expressed as the dilation rate (here expressed as an increase in fault width, Table \ref{tab:1} and Figure \ref{fig:porevolumechange}) times fault slip, divided by the effective fault zone storage. The fault storage consists of the storage capacity of the fault zone $S_\mathrm{f}$ and the width of the fault, so that the undrained isothermal pore pressure drop is
\begin{linenomath}
\begin{equation} \label{eq:pstar}
	\Delta p_\mathrm{undrained} = (dw/d\delta)\delta / S_\mathrm{f}w.
\end{equation}
\end{linenomath}
Undrained conditions may have been approached during dynamic shear failure of intact rock; however, this data cannot be used to estimate fault zone storage since the pore pressure drop was capped by the imposed pore pressure. We can estimate fault zone storage directly from the pore pressure drop measured during the dynamic interval of the stick-slip events, assuming undrained conditions. To do so, we quantified from the data the pore pressure drop $\Delta p_\mathrm{undrained}$ and concurrent fault slip $\delta$ during the dynamic slip events (typically occurring between 1 or 2 data points, i.e., $< 200$~ms). We used the dilation rate measured during progressive slip on the fault, which we approximate as constant regardless of stable sliding or episodes of stick-slip (Figure \ref{fig:porevolumechange}). For a total of 25 analysed stick-slip events measured on 4 samples, most values for $S_\mathrm{f}w$ lie between $1\times10^{-13}$~Pa$^{-1}$m and $6\times10^{-13}$~Pa$^{-1}$m (3 stick-slip events fall below this range, and 3 above). Note that \citet{Brantut2020} followed a similar approach to assess fault zone storage, but a larger time interval was used that includes a partially drained pressure drop and slip preceding dynamic slip. His values therefore must be treated as an upper bound as they may overestimate fault zone storage. 

\begin{figure}
\centering
\includegraphics{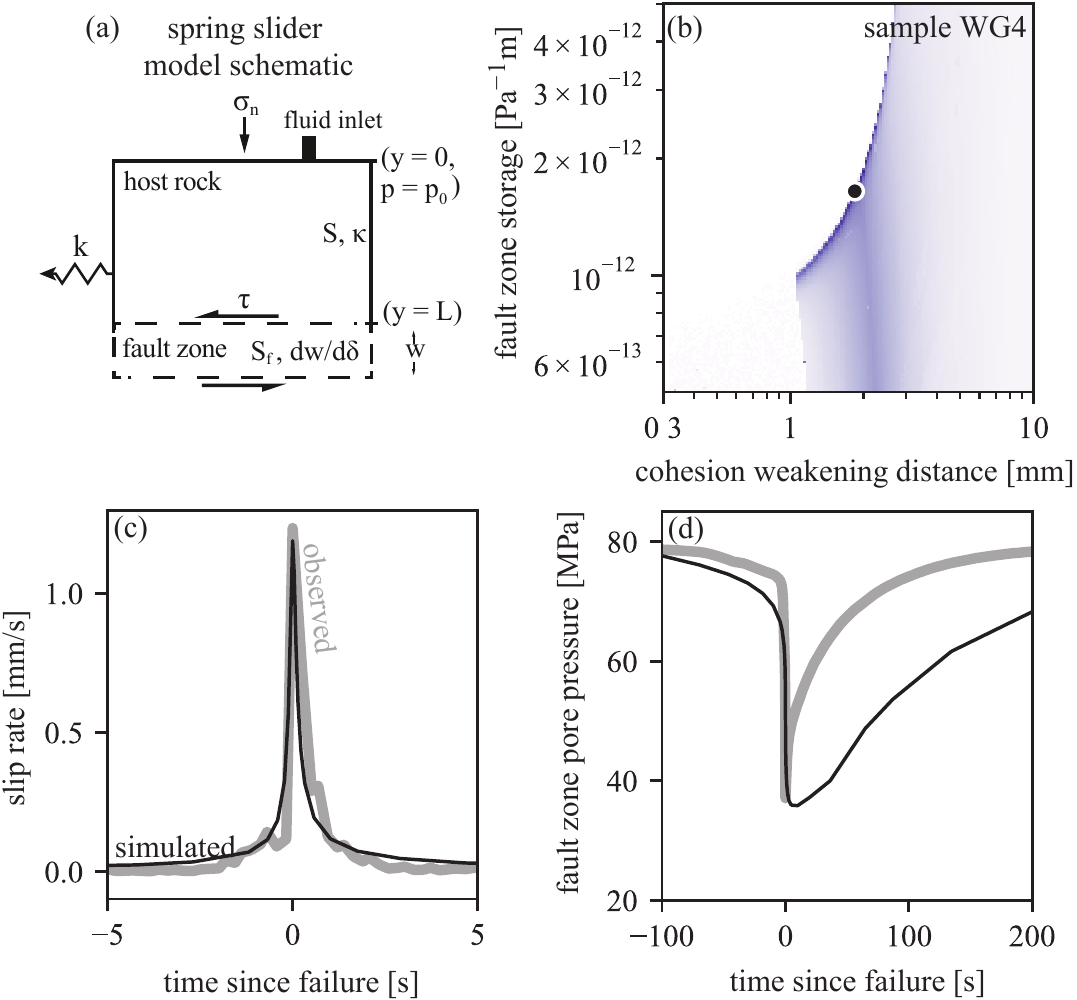} 
\caption{\textbf{(a)}: Sketch of the spring-slider model that allows for dilation within the fault zone. \textbf{(b)}: Probability density resulting from exploring $(\delta_\mathrm{c}, S_\mathrm{f}w)$-space, computed using a least absolute value criterion for the misfit between observed and simulated data for experiment WG4. Best fitting simulation indicated by black marker. \textbf{(c)}: Observed (gray curve) and simulated (black) slip rate over time. \textbf{(d)}: Observed (gray curve) and simulated (black) fault zone pore pressure. }
\label{fig:model}
\end{figure}

During stable shear failure of intact rock the fault zone is partially drained, thus we cannot obtain fault zone storage from equation \eqref{eq:pstar}. Instead, we use a 1D spring-slider model to simulate shear failure under partially drained conditions (Figure \ref{fig:model}a), adapted from \citet{Rudnicki1988}. The spring-slider model uses a cohesive type constitutive law for the strength of the fault, with a residual fault friction that depends on the pore pressure, and allows for fault-normal fluid flow and dilation as a function of slip (see Methods section). We use the spring-slider model and our experimental data in an inverse problem to estimate fault zone storage $S_\mathrm{f}w$ and cohesion-weakening distance $\delta_\mathrm{c}$ for each experiment that exhibited stable shear failure (see Methods section). $\delta_\mathrm{c}$ in the model is akin to the slip-weakening distance measured in triaxial (controlled) rupture experiments on intact granite (Figure \ref{fig:stress-strain}, sample WG6 in Table \ref{tab:1}) \citep[e.g.,][]{Wong1982, Lockner1991, Aben2020a}. However, $\delta_\mathrm{c}$ in the model depends on the definition of cohesion-weakening function and may vary from the measured slip-weakening distances. We therefore treat $\delta_\mathrm{c}$ as an unknown parameter. 

Spring-slider simulations with the best fitting pair of values for $S_\mathrm{f}w$ and $\delta_\mathrm{c}$ simulates fairly well the accelerating slip rate and pore pressure reduction leading up to stable failure (Figure \ref{fig:model}c, d). The simulated on-fault pore fluid pressure recharge after failure does not follow the experimental record (Figure \ref{fig:model}d), likely the result of assuming fault-normal flow only in simulations (further discussed in section 3.2). 

The four experiments yield best fits for $\delta_\mathrm{c}$ that vary between 1.7~mm to 2~mm (Table \ref{tab:1}), similar to values measured for the slip-weakening distance of granite \citep{Aben2020a} (Figure \ref{fig:stress-strain}). We use this to quantify the transition from stable to dynamic failure in our experiments, which occurs when the loading stiffness of the machine $k$ (see Method section for computation of $k$) is lower than the critical stiffness of the fault $k_\mathrm{cr}$. $k_\mathrm{cr}$ depends on the sharpest decrease of the cohesion weakening function ($f(\delta/\delta_\mathrm{c})$ in the spring-slider model, Methods section). For our expression for $f$, $k_\mathrm{cr}^\mathrm{drained} = 1.5\tau_\mathrm{p} / \delta_\mathrm{c}$, where $\tau_\mathrm{p}$ is the cohesion shear stress drop at constant normal stress (Methods section). The undrained critical stiffness is given by $k_\mathrm{cr}^\mathrm{undrained} = (1.5\tau_\mathrm{p} - \mu \Delta p_\mathrm{undrained}) / \delta_\mathrm{c}$. Using representative values for failure at 40~MPa effective pressure ($\tau_\mathrm{p} \approx 85$ MPa, $ \delta_\mathrm{c} \approx 1.5$ mm, $\mu = 0.6$), we see that $k_\mathrm{cr}^\mathrm{drained}/k = 1.16$ and drained conditions always lead to failure. However, the fault is stable (i.e, $k_\mathrm{cr}^\mathrm{undrained}/k < 1$) when the undrained pore pressure drop larger than 28~MPa. This is similar to the smallest partially drained pore pressure drop measured during stable failure (28~MPa at $P_\mathrm{c} = 110, p_0 = 70$, Table \ref{tab:1}). Note that, although the measured partially drained pore pressure drop is not the undrained pore pressure drop, it has the same effect on the critical fault stiffness. As proposed, if the potential pore pressure drop is limited by the magnitude of the imposed pore pressure, the dilatancy strengthening effect cannot reach a sufficient magnitude to stabilise shear failure. Thus we expect vaporisation and dynamic failure if imposed pore pressure $p_0 < 28$ MPa. Indeed, we observe dynamic failure at the experiment performed at $P_\mathrm{c} = 70, p_0 = 30$. Hence, in our experimental setup we always expect stable failure of granite in the presence of pore fluids, provided that the initial pore pressure is sufficiently high to sustain the on-fault pore pressure drop during failure.

The best fitting values for effective fault zone storage $S_\mathrm{f}w$ for the two experiments with the smaller dilation rates are around $1.5\times10^{-12}$~m~Pa$^{-1}$, the two experiments with larger dilation rates yield values that are approximately 3 to 4 times as large (Table \ref{tab:1}). We then simulated the dynamic failure experiments with fault zone storages between $1\times10^{-12}$ and $6\times10^{-12}$ m~Pa$^{-1}$. Using this range, we always predict dynamic shear failure associated with fluid vaporisation (i.e., on-fault pore pressure reaches zero and is capped there in the simulation) for the samples with the lowest imposed pore pressures $p_0=20$ MPa and $p_0=30$ MPa. For dynamic failure at $P_\mathrm{c} = 120$ MPa, $p_0 = 40$ MPa and $P_\mathrm{c} = 160$ MPa, $p_0 = 80$ MPa, dynamic failure was predicted only when $S_\mathrm{f}w$ was smaller than $1.5\times10^{-12}$ and $1.9\times10^{-12}$ m~Pa$^{-1}$, respectively. 

We measured a smaller partially drained pore pressure drops for faults with a larger dilation rate compared to faults with a smaller dilation rate (Table \ref{tab:1}). This may seem counterintuitive at a first glance, but $\Delta p_\mathrm{undrained}$ is effectively the ratio of dilation rate over fault zone storage (equation \eqref{eq:pstar}): The dilation rate doubles from the least dilatant to the most dilatant fault zone, whereas the fault zone storage increases three or fourfold. Of the twin experiments performed at $P_\mathrm{c} = 160$~MPa and $p_0 = 80$~MPa, dynamic failure occurred for a lower dilation rate (and so a larger pore fluid pressure drop leading to vaporisation) compared to the stable failure. 

Effective fault zone storage for stick-slip events is about one order of magnitude smaller than storage estimated for shear failure of intact rock. The difference may be ascribed to i) a change in storage capacity of the fault zone material, which evolves from a micro-fracture dominated zone at the onset of shear failure to a gouge and cataclasite bearing zone of deformation as fault slip progresses, and ii) a difference in the width of the fault zone that is disturbed by dilation. Visual inspection of the samples, and microstructures published by \citet{Brantut2020}, suggest a fault zone width of around 1 to 3~mm for shear failure of intact rock. Microstructures in simulated gouge show strong localisation during stick-slip events, with shear zones of the order of 10 $\mu$m wide \citep{Scuderi2017a}. Dilation outside these principal shear zones is likely, and localisation may be less in rough faults presented here than in simulated gouge. Nonetheless, an order of magnitude decrease in disturbed fault zone width for stick-slip events compared to shear failure remains realistic. 

\subsection{Afterslip driven by pore pressure recovery}
We observed progressive fault slip after stable shear failure that scales linearly with on-fault pore pressure recovery (Figure \ref{fig:recharge}a). As the pore pressure recovers, the shear resistance to faulting decreases by $\tau = \mu (\sigma_\mathrm{n} - p)$. Elastic unloading of the fault's surrounding medium (i.e., the loading piston and the host rock) provides the driving force to overcome the shear resistance, and is given by the elastic unloading stiffness $-k$ of the surrounding medium times fault slip $\delta$. For stable sliding observed in the experiments, shear resistance $\tau$ and imposed load $-k\delta$ are in balance. From this, it follows that afterslip is linearly proportional to pore pressure recovery, with a slope proportional to $\mu / k$. Indeed, the slope of the data can be fitted with this ratio, adjusted for the triaxial conditions of the experiment (Figure \ref{fig:recharge}) (see Methods section for details).

Post-failure slip is primarily driven by diffusive pore pressure recharge of the fault: The fault zone after failure is at a lower pore pressure than its surroundings, and pore pressure reequilibration subsequently occurs at a rate controlled by the hydraulic diffusivity of the fault walls. Specifically, if we assume that the fault is embedded in an infinite  medium (i.e., the distance to any boundary where pore pressure is maintained constant is very large compared to hydraulic diffusion length), the on-fault pore pressure evolution is given by \citep[][, Chap. 12, p. 306]{Carslaw1959}
\begin{equation}\label{eq:trech}
  p(t) -p_0 \approx \Delta p_\mathrm{undrained}\big(1 - e^{t/t_\mathrm{recharge}}\mathrm{erfc}(\sqrt{t/t_\mathrm{recharge}})\big),
\end{equation}
where $t_\mathrm{recharge} = \eta S_\mathrm{f}^2w^2/(4S\kappa)$ is the characteristic recharge time. Using independently measured representative values for each parameter, we obtain $t_\mathrm{recharge}\approx 16$~s, and the time evolution of post-failure pore pressure is well predicted by Equation \ref{eq:trech} (Figure \ref{fig:recharge}b). As stated above and directly observed in the data (Figure \ref{fig:recharge}a), the slip evolution is proportional to the pore pressure evolution, and is also well predicted by the recharge equation \ref{eq:trech} (Figure \ref{fig:recharge}c).

The details of the pore pressure and slip evolution deviate from the simple semi-infinite model because (1) further slip is likely to generate some dilation, limiting the recharge, (2) the imposed constant pore pressure at the ends of the sample accelerates the recharge, and (3) the fault geometry likely leads to along-fault pore pressure diffusion, which also accelerates the recharge, as directly evidenced by the transient time interval where pore pressure is lower off- than on the fault (Figure \ref{fig:porepressure}) -- a situation that is impossible if fluid flow was only one-dimensional. Despite these caveats, the simple recharge model coupled to elastic relaxation explains remarkably well the phenomenon of short-term post-failure slip.

\subsection{Implications for fault slip, earthquakes, and rupture dynamics}
We successfully simulated the experimental findings on dilatancy strengthening using a simple 1D spring slider model, which contains the appropriate ingredients to assess dilatancy strengthening in realistic fault zone settings. The key parameters to assess at which slip rate (or shear rate in a wider zone of deformation) a fault zone transitions from drained to undrained conditions are the hydraulic diffusivity of the host rock outside the zone of dilation, and the fault zone dilation rate and storage. Assessing these latter two parameters for natural fault zones is a challenge for future studies: Larger scale fault zone roughness (e.g., fault bends, dilatational jogs) will influence dilation rate, and the width of the dilating zone (i.e., component $w$ in fault storage $S_\mathrm{f}w$) may vary. In the following section, we discuss the implications of dilatancy strengthening on fault slip, earthquakes, and rupture dynamics using the experimentally constrained parameters, keeping in mind that these values may change according to, amongst others, host rock and fault zone material, and fault roughness. 

\paragraph{Dilatancy-induced pore pressure changes in the crust}
We shall first predict the dilatancy-induced pore pressure drop in the crust for materials with large cohesion, such as intact rock and consolidated faults, for the two end-member cases of fully drained and fully undrained conditions. The dilation rate during shear failure of intact rock decreases with increasing effective pressure, as we can see from our data and those of \citet[][Figure 6]{Rummel1978}, which can be described by an exponential function (Figure \ref{fig:Rummel_fit}). We used this function to express a mean dilation rate with upper- and lower bounds (Figure \ref{fig:Rummel_fit}). With these values, the undrained isothermal pore pressure drop (equation \eqref{eq:pstar}) is between 50 and 100~MPa at effective pressures below 50 MPa, and becomes negligible at effective pressures above 200~MPa (Figure \ref{fig:pstar}a). 

\begin{figure}
\centering
\includegraphics{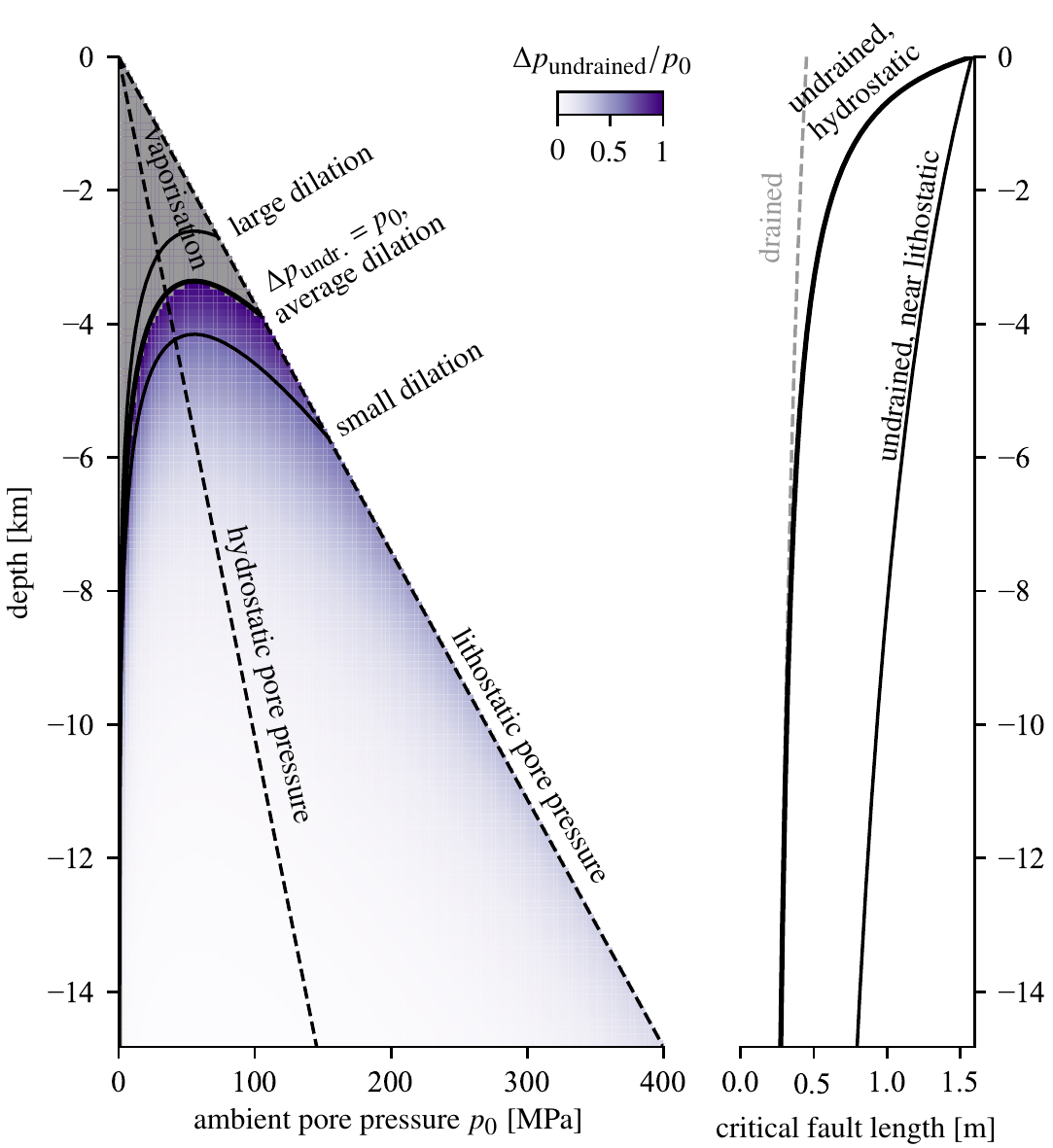}
\caption{Undrained isothermal pore pressure drop during shear failure of intact material in a crustal depth profile with a range of ambient pore pressures (hydrostatic and lithostatic pore pressure profiles shown as dashed lines). The undrained pore pressure drop is normalised by the ambient pore pressure, shown here for average fault zone dilation. Black curves show where the undrained pore pressure drop equals the ambient pore pressure (i.e., net zero pore pressure), for average, small, and large dilation. At depths smaller than these curves, vaporisation is expected. Left panel: Critical length of the slipping patch of the fault, below which the fault is stable. Gray curve: drained critical fault length, black curves: undrained critical fault length for a hydrostatic pore pressure profile and a near lithostatic pore pressure profile. Critical fault length calculated with a shear modulus of 25~GPa and poisson's ratio of 0.22. }
\label{fig:crust}
\end{figure}

Effective pressure in the crust depends on the depth and the ambient pore fluid pressure, which typically varies between hydrostatic and lithostatic pore pressure. Using the pressure-dependent undrained pore pressure drop, we predict that vaporisation during shear failure is likely to occur down to a depth of 4~km (for average dilation) regardless of the initial pore pressure (Figure \ref{fig:crust}). At very low pore pressure, vaporisation may occur down to 8~km depth. We thus do not expect significant failure stabilisation in the uppermost part of the crust. We do however expect a strong dilatancy strengthening effect in the 2 kilometres directly below the vaporisation zone, where the undrained pore pressure drop is between 20\% and 100\% of the ambient pore pressure. Dilatancy strengthening becomes negligible at high effective pressures, below around 6 km depth, but remains significant in regions in the crust with a low effective pressure -- i.e., regions where the pore pressure approaches lithostatic pressure. 

In the above analyses, the fault zone storage $S_\mathrm{f}w$ remained constant with effective pressure. Storage capacities of rocks typically decrease with increasing pressure (see for instance Figure \ref{fig:permstorativity}), and we expect the same for the fault zone storage capacity $S_\mathrm{f}$. Qualitatively, a pressure-dependent fault zone storage capacity would yield a larger undrained pore pressure drop at higher effective pressure, i.e., greater depth. 

Note that our estimates of pressure drop rely on a slip-dependent dilation model, which is representative of slip occurring on intact or consolidated faults, possibly with some natural roughness. The direct slip dependence is motivated by observations (Figure \ref{fig:porevolumechange}), but also reflects the end-member case of a rate-and-state fault subject to rapid steps in slip rate without healing. In that context, the evolution of porosity with slip reported here reflects irreversible state evolution of the fault. For mature, smooth faults containing unconsolidated fault gouge (relevant at shallow depth), a model where dilatancy and compaction fully depends on rate and state evolution \citep{Marone1990, Samuelson2009} would be more appropriate, although this dependence is rather small compared to the dilation measured here \citep[][Section 5.3]{Brantut2020}.

\paragraph{Implications for fault slip}
We predict strong dilatancy strengthening in the upper crust (between 4 and 6 km) and at increased depth if the ambient pore pressure approaches lithostatic pressure. This has implications for the behaviour of fault slip and rupture dynamics across the entire ``spectrum of fault slip'' from slow slip rates to earthquakes. 

In the last two decades, a range of fault slip rates in different crustal settings have been measured that are slower than dynamic slip during earthquakes \citep{Peng2010, Burgmann2018}, such as slow slip events (SSEs) at the base of the seismogenic crust in subduction zones and in crustal-scale strike slip faults. Ambient pore pressure in these regions are typically high and may approach the lithostatic pore pressure \citep[e.g.,][]{Thomas2009, Matsubara2009}. Many SSEs are episodic, and may occur on parts of the fault that are also subject to fast slip (i.e., earthquakes) \citep{Burgmann2018}. Slow slip events require initial weakening of the fault zone material to achieve a notable increase of fault slip, followed by some strengthening mechanism to limit the slip rate and rupture velocity. Dilatancy strengthening of a velocity-weakening material has been proposed as one of these strengthening mechanisms \citep[e.g.,][]{Segall2010}, motivated by the abundance of pore fluids within most SSE settings. Our experiments support this idea directly, observing that dilatancy strengthening leads to the formation of an 8~cm long fault in around 3~s, giving a propagation velocity of around 2.3~km per day that matches with propagation velocities for SSEs of around 10~km per day \citep{Gomberg2016}.

We observed that dilatancy generates two separate phases of fault slip: The initial weakening process followed by prolonged fault slip and stress reduction driven by pore pressure recharge of the fault. The timescale of the first phase is an elastodynamic one, governed by the slip-weakening distance and, in our tests, machine stiffness, and in nature by elastic properties of the surrounding rock. The rate of the second phase is determined by the characteristic timescale $t_\mathrm{recharge}$ (section 3.2, equation \ref{eq:trech}), which depends on dilation zone width $w$. If the recharge timescale is much larger than the initial elastodynamic weakening timescale, pore pressure reequilibration can lead to prolonged slip far from the rupture tip within a single rupture event. The characteristic timescale for the experiments is around 16~s. However, in nature, the dilatant region could be orders of magnitude larger due to fault roughness. In this case, afterslip may be observed as transient post-seismic slip, with post-seismic fluid migration potentially driving aftershock sequences \citep[e.g.,][]{Nur1972a, Miller2020}. 

\paragraph{Implications for earthquake nucleation}
Strong dilatancy strengthening may not always prevent earthquake nucleation. An earthquake nucleates when the loading stiffness of the material surrounding the fault is lower than the faults' critical stiffness ($k_\mathrm{cr}$). The loading stiffness decreases with growth of the slipping fault section, whilst the faults' critical stiffness decreases with dilatancy as follows: We use the same formulation for drained and undrained $k_\mathrm{cr}$ as in section 3.1, now with pressure-dependent values for $\Delta p_\mathrm{undrained}$ and $\tau_\mathrm{p}$. Pressure-dependent $\tau_\mathrm{p}$ for Westerly granite was obtained from \citet[][Figure 5 and 9]{Byerlee1967}. At effective pressures below 200~MPa, the undrained critical stiffness is greatly reduced compared with the drained critical stiffness (Figure \ref{fig:pstar}b). At higher effective pressures, the dilatancy strengthening effect on the critical stiffness vanishes. The loading stiffness of the surrounding material for faults embedded in an elastic continuum is approximated by $k=G/W$, where $G$ is the shear modulus and $W$ is the length of the slipping fault section. Therefore, the critical stiffness is inversely proportional to a critical length of the slipping fault section, so that the slipping fault section becomes unstable when it exceeds this critical nucleation length. Using our data on Westerly Granite, the drained nucleation length remains more or less stable throughout the brittle crust at around 0.3 to 0.4~m (Figure \ref{fig:crust}). For an undrained fault zone at ambient hydrostatic pore pressure, the nucleation length in the top two km of the crust increase fourfold (i.e., in the zone where we predict vaporisation), and reverts to the drained nucleation length at increased depth. For undrained conditions at ambient pore pressures approaching lithostatic pore pressure, we expect that the nucleation length increases by a factor of 4 below the vaporisation zone and by a factor of 3 at 15 km depth (Figure \ref{fig:crust}). 

The near-lithostatic pore pressure zones correspond to the zones where SSEs occur, and although the nucleation length of Westerly granite may not be representative for the material that host SSEs, we still expect a strong increase in nucleation length for these materials. The critical nucleation length for Westerly granite is directly applicable in regions in the crust with human subsurface activities, such as geothermal energy reservoirs. These are generally located between 2 and 5~km depth \citep[e.g.,][]{Tomac2018}, which overlaps with the depth range for pore fluid vaporisation and for strong dilatancy hardening (Figure \ref{fig:crust}). 

Dilatancy strengthening leads to slow slip over increased distances, but does not necessarily change the inherent seismogenic character of the material and also allows for seismicity. This is evident from the stick-slip events during continued fault slip in our experiments, where dilatancy strengthening was insufficient to suppress dynamic rupture. The approach towards dynamic rupture is however extended by dilatancy strengthening. Such a longer precursory phase may allow for better identification of active earthquake precursory phenomena such as foreshocks \citep[e.g.,][]{Ohnaka1992, Bouchon2013} or precursory creep \citep[e.g.,][]{Roeloffs2006}, and passive phenomena such as changes in $v_\mathrm{p} / v_\mathrm{s}$ ratios \citep[e.g.,][]{Nur1972} and changes in seismic wave amplitudes \citep[e.g.,][]{Shreedharan2021}. These two latter phenomena may be particularly sensitive to local pressure changes and fracture damage induced by dilatancy \citep{Shreedharan2021}. Hence, crustal regions with strong dilatancy strengthening may be best suitable to find reliable precursory phenomena. 

\paragraph{Implications for rupture energy budget}
Dilatancy strengthening affects the dynamics of rupture. Rupture propagation is governed by the partitioning of stored elastic strain energy: A small part of it may be released as radiated energy when failure is dynamic, whereas most is dissipated to overcome residual fault friction during sliding (frictional work) and as breakdown work done in excess of the frictional work. Breakdown work $W_\mathrm{b}$ is a collective term for energy spent to reduce the intact strength of the fault zone towards its residual steady-state frictional strength. The work done to overcome the dilatancy-induced extraneous shear resistance that we observe in our experiments is in excess of the (drained) residual friction, and is thus part of the breakdown work. Breakdown work can be estimated simply as the area under the shear stress versus slip curve in excess of the residual shear stress \citep{Wong1982} (Figure \ref{fig:stress-strain}b). Doing so for the experimental data provides values for $W_\mathrm{b}$ at stable shear failure of 102~kJm$^{-2}$ (sample WG12), 104~kJm$^{-2}$ (sample WG5), and 122~kJm$^{-2}$ (sample WG4) at 40~MPa effective pressure, and 143~kJm$^{-2}$ at 80~MPa effective pressure (sample WG8). The minimum breakdown work necessary to form a fault zone and reach the residual frictional strength of a rock (also known as shear fracture energy) may be measured from a quasi-static or ``controlled'' shear failure experiment \citep[e.g.,][]{Lockner1991, Aben2020a}. We performed such experiments on a notched, thermally cracked Westerly granite samples at 40~MPa effective pressure (Figure \ref{fig:stress-strain}b), and obtain $W_\mathrm{b} = 76$ kJm$^{-2}$. For slow shear rupture, we thus see a 34-60\% increase in breakdown work attributed to dilatancy strengthening at 40~MPa effective pressure, and so less released strain energy may be used to accelerate rupture and slip -- the failure process remains stable. 

We expect a further increase in breakdown work as rupture accelerates. At low rupture velocity, we have shown that the dilatancy-induced pore pressure drop near the rupture tip is partially drained, as the measured pore pressure drop is less than $\Delta p_\mathrm{undrained}$. At higher velocity, the fault zone dilates in a shorter time interval and so the pore pressure drop approaches the undrained pore pressure drop. The resistance to slip thus increases more for a fast rupture than for a slow rupture. The strengthening effect remains transient and vanishes with pore pressure recharge, but the time delay, or distance along the fault, between the initial stress drop at the rupture tip and the second stress drop from pore pressure recovery is controlled by how fast the rupture propagates and how fast pore pressure diffusion can compensate the undrained pressure drop near the tip. Thus, the breakdown work increases with rupture velocity 1) because of a larger dilatancy strengthening effect, and 2) because a larger amount of slip is accumulated before the fault zone pore pressure reaches its ambient value. 

Weakening processes acting behind the rupture tip at faster fault slip may be impacted by dilatancy strengthening. For example, we expect that the onset of weakening by thermal pressurisation of pore fluids \citep[e.g.,][]{Lachenbruch1980, Rice2006} will be delayed by the dilatancy-induced pore pressure drop at the rupture tip compared to a fully drained case, as the thermal pressurisation process needs to overcome the deficit in pore pressure first. This may increase the temperature in the fault zone due to frictional heating, which is otherwise buffered by thermal pressurisation weakening \citep{Garagash2003}.

\section{Conclusions}
Our laboratory experiments demonstrate that dilatant strengthening stabilises rock failure and fault slip. The effect of dilatancy is capped by the zero lower bound for pore pressure, where fluid vaporises. All our tests where failure was dynamic experienced transient fluid vaporisation. In the presence of pressurised fluids, rupture occurs in two stages: An initial stage driven by intrinsic weakening and elastic energy release form the surrounding medium, and a second stage where post-failure pore pressure reequilibration leads to prolonged slip and stress drop, purely controlled by pore pressure changes. Our laboratory data are quantitatively explained by a simple spring-slider model, which we use to constrain a key previously unknown quantity, the fault zone storage capacity. The consequences of dilatant stabilisation of rupture are manifold, including an increase in nucleation size, slowing of rupture propagation and increases in breakdown work. Our laboratory techniques opens the way to systematic quantification of hydro-mechanical parameters under in-situ conditions, so that the wealth of theoretical knowledge on diltancy \citep[e.g.,][]{Rice1979b,Segall1995,Segall2010,Ciardo2019} can be used and testable predictions can be made.

\section*{Acknowledgements}
This study was funded by the UK Natural Environmental Research Council grant NE/S000852/1 to N.B., and the European Research Council under the European Union's Horizon 2020 research and innovation programme (project RockDEaF, grant agreement \#804685 to N.B.). We thank Neil Hughes for technical assistance with development of the pore pressure sensors. All data needed to evaluate the conclusions in the paper can be found at the NGDC repository of the British Geological Survey (https://www.bgs.ac.uk/services/NGDC, dataset ID165485). 

%\bibliographystyle{agufull08}
%\bibliography{litall.bib}

\clearpage

\section*{Methods}
\paragraph{Sample preparation}
Cylindrical Westerly granite samples of 40~mm diameter and 100~mm height were cored and their surfaces were ground parallel. We cut two 17~mm deep notches into the cylindrical surface at a $30^{\circ}$ angle with the cylinder axis (Figure \ref{fig:A}a). The notches were aligned opposite each other so that the plane in between was most likely to fail during axial loading. The samples were subjected to thermal microcracking to increase the hydraulic diffusivity of the rock, achieved by placing the samples in a tube furnace that was heated at a rate of $3^{\circ}$C~min$^{-1}$ to $600^{\circ}$C. This temperature was maintained for the duration of two hours, followed by cooling over the course of about 12 hours by switching off the furnace. The notches were filled with Teflon disks prior to insertion in a rubber jacket (Figure \ref{fig:A}a). The Teflon spacers do not compact or dilate, they deform in a plastic manner, have a low frictional resistance to sliding, and are considerably weaker than Westerly granite. The plastic behaviour is evidenced by the shape of the spacers at the end of each experiment. Hence, we assume that the spacers do not impact the mechanical behaviour of the fault by pinning the fault plane or inhibiting fault sliding. The jacket was equipped with four miniature pore pressure sensors: Two on the prospective failure plane, and two at the same height on the intact part  of the sample. 

\paragraph{Pore pressure sensors}
The pore pressure sensors consist of a steel 12~mm diameter cap with a thickness of 2.5~mm (Figure \ref{fig:A}b). On the inside edge of the cap, an 0.2~mm high lip creates a small reservoir in which pore fluid resides. The cap is placed over a metal stem so that the inside lip rests on the stem's surface. The pore fluid within the cap is isolated from the confining medium by an O-ring on the stem. The metal stem is glued into the rubber jacket, and allows pore fluid to pass from the metal cap to the surface of the sample through an 0.4~mm diameter wide bore in its centre. A diaphragm strain gauge (four individual strain gauges arranged in a circular pattern, where two strain gauges measure tangential strain and two measure circumferential strain, wired in a full bridge configuration) is bonded to the outer surface of the steel cap. When pressure is applied to the steel cap from the outside (i.e., by confining pressure), the steel cap will elastically deflect inwards. Pore fluid pressure applied from the inside of the steel cap will induce elastic deflection in the opposite direction. The elastic strain of the steel cap with effective pressure (confining pressure minus pore fluid pressure) is measured as a linear change in resistivity of the diaphragm strain gauge. The change in resistivity was measured by a Fylde DC Transducer and Amplifier and digitally logged. The sensor were calibrated prior to the onset of deformation by a number of confining pressure steps and pore pressure steps. Further details on these sensors and their calibration are found in \cite{Brantut2021}. 

\paragraph{Experimental setup}
The shear failure experiments were executed in a conventional oil-medium triaxial loading apparatus at University College London. Axial load was measured by an external load cell corrected for friction at the piston seal. Axial shortening was measured by a pair of Linear Variable Differential Transducers (LVDTs) outside the confining pressure vessel, corrected for the elastic shortening of the piston. Up- and downstream pore pressures were measured by pressure transducers located outside the pressure vessel, and pore pressure was controlled by a single pore pressure intensifier equipped with an LVDT to measure the volume change in the intensifier reservoir. 

The samples were deformed at nominal effective pressures of 40 MPa or 80 MPa, with a varying combination of confining and imposed pore pressure (Table \ref{tab:1}). The imposed pore pressure during deformation was kept constant at both sample ends. Axial load was increased by imposing a constant piston displacement rate of $1\times 10^{-4}$~mm s$^{-1}$ that corresponds to a strain rate of $1\times 10^{-6}$~s$^{-1}$. The samples were loaded until shear failure, after which the piston displacement was arrested to measure the pore volume change $\Delta V_\mathrm{p}$ in the sample. After the measurements, we continued to accumulate slip on the newly formed fault; either by stable sliding or by stick-slip events. After each slip increment, we paused deformation to remeasure pore volume changes. Pore volume change was measured from the volume change in the pore pressure intensifier, which we ascribe entirely to the change in pore volume in the sample. After arresting the movement of the loading piston, the pore pressure throughout the sample was allowed to recover towards the imposed value. This recovery was achieved when the volume of the pore pressure intensifier reached a stable value. We defined the pore volume change for an interval of fault slip between two such equilibration points. Note that we expect some minor compaction of fault material when waiting for equilibration of the pore volume between slip intervals, so that the actual fault zone porosity during sliding might be slightly higher. The first point in this series is at zero fault slip, defined at the peak differential stress where we did not arrest deformation, but nonetheless recorded a nearly homogenous pore pressure throughout the sample.

\paragraph{Corrections for normal and shear stress}
The normal and shear stress on the fault are both a function of differential stress and confining pressure and where calculated following equation \eqref{eq:triax} with a known angle $\psi = 30^{\circ}$. Two corrections should then be applied to the normal and shear stresses: One for the reduction in fault contact area as a function of fault slip, and one for the presence of the Teflon spacers. We did not apply the latter correction to the normal stress record, as we assume that the spacers support normal stresses, but we did correct the normal stress for the reduction in fault contact area with fault slip using equations A3 and A4 from \cite{Tembe2010}. 

The shear stress was corrected for the presence of the Teflon spacers, which have a lower shear resistance than the rock. We assume a constant coefficient of friction of the rock-Teflon-rock ``sandwich'' $\mu_\mathrm{teflon}$, so that the maximum shear force supported by the spacers is:
\begin{linenomath}
\begin{equation}
F_\mathrm{teflon} = \sigma_\mathrm{n} \mu_\mathrm{teflon} \times 2A_\mathrm{teflon},
\end{equation}
\end{linenomath}
with normal stress $\sigma_\mathrm{n}$ (already corrected for contact area reduction), and the surface area of the spacer $A_\mathrm{teflon}$. The total shear force is given by the uncorrected shear stress times the total fault surface area. We then obtain the corrected shear stress as the difference between the total shear force and $F_\mathrm{teflon}$, divided by the difference between total surface area and $2A_\mathrm{teflon}$. However, we do not have direct measurements for the coefficient of friction $\mu_\mathrm{teflon}$. We therefore estimate $\mu_\mathrm{teflon}$ by fitting the corrected peak shear stress with the peak shear stress measured on samples of Westerly granite without notches at 50 and 100 MPa confining pressure (unpublished data, Figure \ref{fig:shearcorrection}). Both sets of data were obtained using the same triaxial deformation apparatus, had the same sample dimensions, and samples were cored from the same blocks of Westerly granite. We could match the trend of peak shear stress versus normal stress of the unnotched samples for a shear stress correction with $\mu_\mathrm{teflon} = 0.4$ (Figure \ref{fig:shearcorrection}). This shear stress correction was applied to all shear stress data in this study. We neglected the shear stress correction for the reduction in fault contact area because the contact area that is lost consists of Teflon, resulting in a small correction factor that is of the same order of magnitude as the uncertainties in estimating the correction for the presence of the Teflon spacers. 

\paragraph{Calculation of dilation rate}
We compute dilation rate from the pore volume data, assuming any change in pore volume occurred in the fault zone after the peak stress was surpasses. Similar to shear experiments on simulated gouge in a direct shear setup, our fault zone experiences i) simple shear, ii) dilation or compaction of the fault zone material, and iii) extrusion of material by fault sliding at the sample edges leading to geometrical thinning \citep[e.g., Figure 6 in][]{Scott1994}. The pore volume change we measure is thus the volume change by ii) plus the volume change by iii). We note that at the start of our experiment the fault zone has its lowest porosity because the prospective fault plane consists of intact granite. Hence, in the knowledge that Teflon does not dilate or compact (i.e., the volume of the spacers remains constant), the Teflon spacers do not prop open the fault zone, nor do they contribute to changes in pore volume by contribution ii). However, as the Teflon spacers are positioned at the fault extremities, they will extrude somewhat, leading to geometrical thinning and reduction of fault zone volume. 

We express dilation as the rate of increase in fault zone width $w$ with fault slip $\delta$. The volume of the fault zone is approximated as an elliptic cylinder with fault zone width $w$, a long axis radius $a = 40$~mm and a short axis radius $b = 20$~mm. The axes remains constant during deformation so that all pore volume change that has been measured is accommodated by a change in $w$. This allows us to calculate $d w = \Delta V_\mathrm{p} / \pi a b$ for the slip interval in which intact failure occurred, and for the slip intervals after shear failure. Fault slip was calculated from the 30$^\circ$ angle between the fault and the direction of axial load and the axial strain measurements corrected by the intact Young's modulus of the sample, thereby assuming that all deformation after the peak stress was accommodated by fault slip. The elliptic cylinder, which we assume hosts all pore volume change, partly consists of volume-neutral Teflon spacers, and so projecting the pore volume change in an elliptic cylinder leads to an underestimation of the fault zone porosity change. This underestimate is offset by extrusion of the Teflon spacers from the fault zone and by some pore volume change due to damage outside the main failure zone observed on the post-mortem samples. 

Here, dilation rate has been presented as a function of fault slip $\delta$, which is the desired representation of the dilation rate for slip-weakening or cohesion-weakening models that include dilatancy strengthening \citep[e.g.,][]{Rudnicki1988}. Dilation has also been expressed as a function of an increase in imposed slip rate for frictional sliding experiments on simulated gouge in the context of rate and state friction \citep{Marone1990, Samuelson2009}. In these studies, the overall trend of gouge compaction from geometric thinning was removed from the gouge thickness data to reveal small changes related to steps in slip rate. The increase in fault zone width $w$ was directly measured and normalised by the initial width of the gouge layer. We cannot compare our data with these studies because: i) Slip rate steps were not performed as we arrested slip in between slip intervals to measure pore volume changes. ii) Our setup cannot (yet) continuously measure the instantaneous pore volume increase from monitoring the fault zone thickness directly. iii) We do not know the initial thickness of the fault zone required to normalise dilatancy as defined by \citet{Marone1990, Samuelson2009}.

\paragraph{Hydraulic characterisation of intact rock}
Permeability $\kappa$ and storage capacity $S$ of the intact rock were obtained prior to deformation by a transient pore pressure front method akin to the pulse-decay method. A 5~MPa pore pressure step was produced at a rate of 5~MPa s$^{-1}$ in the upstream pore pressure intensifier. The downstream end of the sample was undrained, with a known downstream storage capacity. The pore pressure records and the upstream pore volume change for the pore pressure pulse may be expressed by a closed-form solution with two unknown parameters that are commensurate to permeability and storage capacity. The solution can be used for any location along the sample height between the upstream and downstream reservoirs, assuming homogenous hydraulic properties in the sample. This allows us to invert the pore pressure and pore pressure intensifier volume records to obtain the best-fitting pair of values for the permeability and storage capacity, with the intermediate pore pressure records measured by the effective pressure transducers providing some additional constraints to the solution. For more details on the pore pressure front method, see \citet{Brantut2021}. The results of the hydraulic characterisation of the samples are presented in Figure \ref{fig:permstorativity}.

\paragraph{Spring-slider model: Setup}
We consider the sample as a rigid body split by a shear fault (Figure \ref{fig:model}a). We impose a fault slip rate $v_\infty$ through an elastic medium with stiffness $k$ in the direction of fault slip $\delta$. For quasistatic fault motion, the imposed load is equal to the shear resistance $\tau$ of the fault:
\begin{linenomath}
\begin{equation}\label{eq:eq}
k(v_\infty t - \delta) - \tau -\chi v= 0,
\end{equation}
\end{linenomath}
where $v$ is the slip rate, and $\chi$ is a viscous damping parameter preventing velocity to become unbounded during instabilities. Typically $\chi v$ remains negligible compared to other terms unless $v$ becomes extremely large \citep[e.g.,][]{Segall1995}. We then follow \citet{Rudnicki1988} and use a cohesive type constitutive law for the strength of the fault, with a residual fault friction that depends on the pore pressure (as seen in our data, Figure \ref{fig:recharge}):
\begin{linenomath}
\begin{equation}\label{eq:tau}
\tau = \underbrace{\tau_\mathrm{p} (1-f(\delta/\delta_\mathrm{c}))}_\text{cohesion} + \underbrace{\mu(\sigma_\mathrm{n}-p)}_\text{friction},
\end{equation}
\end{linenomath}
where $\tau_\mathrm{p}$ is the drop in shear stress from fracture strength to frictional strength for a constant normal stress (Figure \ref{fig:ss_sketch}). The function $f$ describes the loss of cohesion with slip, with $f(0) = 0, f(1) = 1$. $\delta_\mathrm{c}$ is a characteristic slip-weakening distance (Figure \ref{fig:ss_sketch}a). Here, we use $f = -2 (\delta/\delta_\mathrm{c})^3 + 3(\delta/\delta_\mathrm{c})^2$ as suggested by \citet[][equation 3]{Rudnicki1988} (Figure \ref{fig:ss_sketch}a). The shape of this cohesion-weakening function matches well with the overall shape of the quasi-static controlled rupture (Figure \ref{fig:stress-strain}a). 

To simulate changes in $p$, we first consider that the fault zone itself is of a uniform width $w$ and has a uniform pore pressure, so that only fault-normal fluid flow may occur in spirit with the 1D spring-slider system. The fault zone is considered a ``reservoir'' with effective fault zone storage $S_\mathrm{f}w$. The fault zone is in hydraulic communication with the intact host rock, so that the pore pressure is governed by a diffusion equation:
\begin{linenomath}
\begin{equation}\label{eq:poff}
  \frac{\partial p}{\partial t} = \frac{\kappa}{\eta S} \frac{\partial^2 p}{\partial y^2},
\end{equation}
\end{linenomath}
where $\kappa$ is host rock permeability, $S$ is the storage capacity of the host rock, and $\eta$ is the fluid viscosity. The coordinate $y$ is the fault-normal spatial dimension, with $y=0$ being the upstream end of the sample, and $y=L$ the fault position (i.e., the half-length of the sample) (Figure \ref{fig:model}a). The experimental boundary conditions impose that $p(y=0,t)=p_0$. Note that the system is symmetric here so we only consider one half of the sample. A mass balance leads to the following boundary condition at $y=L$:
\begin{linenomath}
\begin{equation}\label{eq:pon}
\frac{\partial p}{\partial t} + \frac{2\kappa}{S_\mathrm{f} w \eta}\frac{\partial p}{\partial y} = -\frac{\dot{\Delta \phi}}{S_\mathrm{f}} \quad (y=L),
\end{equation}
\end{linenomath}
where $\dot{\Delta \phi}$ is the rate of inelastic porosity change inside the fault zone -- describing dilation as a function of slip. Recall that from the experimental data, we expressed dilation as a linear increase in fault zone width with slip $d w / d\delta$. We then obtain the rate of porosity change as $\dot{\Delta \phi} = (d w / d\delta)/w \times v$. For undrained conditions, equation \eqref{eq:pon} reverts back to equation \eqref{eq:pstar} for the undrained isothermal pore pressure drop. Further details to account for triaxial conditions, normalisation of variables, and numerical implementation to solve the problem are presented in the next paragraph.

\paragraph{Spring-slider model: Application to triaxial experiments}
The spring-slider model as defined by equations \eqref{eq:eq} and \eqref{eq:tau} are for a constant normal stress and a spring load parallel to the sliding direction. Here, the experiments were conducted under triaxial conditions, so that the normal and shear stress are both a function of the differential stress $Q$ and confining pressure $P_\mathrm{c}$:
\begin{linenomath}
\begin{align} \label{eq:triax}
  \tau &= (Q/2)\sin(2\psi),\\
  \sigma_\mathrm{n} &= P_\mathrm{c} + (Q/2)\big(1-\cos(2\psi)\big),
\end{align}
\end{linenomath}
where $\psi$ is the angle between loading direction and the fault zone. From this, we obtain
\begin{linenomath}
\begin{equation}
\tau = \tau_0 (1-f(\delta/\delta_\mathrm{c})) + \mu(P_\mathrm{c} - p) / \left(1-\mu\frac{1-\cos(2\psi)}{\sin(2\psi)}\right),
\end{equation}
\end{linenomath}
with
\begin{linenomath}
\begin{equation} \label{eq:tau0}
\tau_0 = \tau_\mathrm{p} / \left(1-\mu\frac{1-\cos(2\psi)}{\sin(2\psi)}\right).
\end{equation}
\end{linenomath}
We understand $\tau_0$ to be the shear stress drop measured in triaxial failure experiments (Figure \ref{fig:ss_sketch}b). The loading piston and intact parts of the rock sample act as the spring load. Differential stress decreases from elastic relaxation of the loading piston by $\Delta Q = -k' \epsilon$, with $k'$ being the machine stiffness and $\epsilon$ the axial elastic lengthening of the piston. As $Q$ decreases, the rock sample increases in length by elastic rebound as $\epsilon k' / E \times 2L$, where $L$ is the half length of the sample and $E$ is the Young's modulus of the sample. Because the movement of the top of the piston was arrested during shear failure, $\epsilon$ and the relaxation of the rock sample are both accommodated by slip along the fault $\delta$ so that $\epsilon = \delta \cos(\psi) / (1+2k' E/L)$. From this and equation \ref{eq:triax}, it follows that the fault parallel stiffness $k$ is
\begin{linenomath}
\begin{equation} \label{eq:machinestiffness}
k = \frac{k' \sin(2\psi)\cos(\psi)}{2 (1+2 k' L/E)}. 
\end{equation}
\end{linenomath}

\paragraph{Spring-slider model: Numerical solution for partially drained case}
We nondimensionalise our governing equations prior to numerical implementation for partially drained conditions. We use the following scales: 
\begin{linenomath}
\begin{align}
  t\leftarrow t/t_\mathrm{diff},\\
  y\leftarrow y/L,\\
  p\leftarrow p/\Delta p_\mathrm{undrained},\\
  \tau\leftarrow \tau/\tau_0\\
  \delta\leftarrow \delta/\delta_\mathrm{c},\\
  v\leftarrow v/(\delta_\mathrm{c}/t_\mathrm{diff}),
\end{align}
\end{linenomath}
where
\begin{linenomath}
\begin{equation}
  t_\mathrm{diff} = \frac{L^2 \eta S}{\kappa}
\end{equation}
\end{linenomath}
is the diffusion timescale across the sample, the undrained isothermal pore pressure drop $\Delta p_\mathrm{undrained}$ is given by equation \eqref{eq:pstar} (Figure \ref{fig:ss_sketch}b) (with $d w / d\delta$ normalised w.r.t. $\delta_\mathrm{c}$ already), and $\tau_0$ is given by equation \eqref{eq:tau0}. The static equilibrium (equation \eqref{eq:eq}) is thus rewritten as
\begin{linenomath}
\begin{equation}\label{eq:eqnorm}
  K(v_\infty t- \delta) - (1-f(\delta)) - \tau_\mathrm{D}(\lambda - p) -X v= 0,
\end{equation}
\end{linenomath}
where
\begin{linenomath}
\begin{align}
  K &= k \delta_\mathrm{c}/\tau_0,\\
  \tau_\mathrm{D} &= \mu \Delta p_\mathrm{undrained}/\tau_\mathrm{p},\\
  \lambda &= P_\mathrm{c}/\Delta p_\mathrm{undrained},\\
  X &= \chi \delta_\mathrm{c}\kappa/(L^2\tau_0 \eta S).
\end{align}
\end{linenomath}
Note that $\tau_\mathrm{D}$ is to be understood as the cohesion loss over the undrained pore pressure change times the friction coefficient -- i.e., the undrained dilatancy-induced increase in shear resistance (Figure \ref{fig:ss_sketch}b). The governing equation for pore pressure becomes
\begin{linenomath}
\begin{equation} \label{eq:pdiffnorm}
  \frac{\partial p}{\partial t} = \frac{\partial^2 p}{\partial y^2},
\end{equation}
\end{linenomath}
with boundary conditions
\begin{linenomath}
\begin{align}
  p(0) &= p_0/\Delta p_\mathrm{undrained},\label{eq:BCtop}\\
  \displaystyle \frac{\partial p}{\partial t} + \ell\frac{\partial p}{\partial y} &= -v \quad (y=L),\label{eq:BCfault}
\end{align}
\end{linenomath}
where
\begin{linenomath}
\begin{equation}
  \ell = \frac{2SL}{S_\mathrm{f}w}.
\end{equation}
\end{linenomath}

The equation we are solving numerically is simply:
\begin{linenomath}
\begin{align}
   \frac{dv}{dt} &= \left[K(v_\infty - v) - f'[\delta]v + \tau_\mathrm{D}\dot{p}\right]/\chi, \label{eq:ode1}\\
  \frac{d\delta}{dt} &= v \label{eq:ode2}, 
\end{align}
\end{linenomath}
where we compute the pressure rate in the fault by solving the full diffusion problem (Equation \ref{eq:pdiffnorm}) using finite differences in space and the method of lines, so we can use very efficient stiff ODE solvers for the task. Relabelling the space coordinate $y=1-y$, so that the fault is at position $y=0$ and sample edge is at $y=1$, the diffusion problem reads
\begin{linenomath}
\begin{align}
  \frac{\partial p}{\partial t} &= \frac{\partial^2 p}{\partial y^2},\\
  p(y=1,t) &= 1,\\
  \frac{\partial p}{\partial t}\Big|_{y=0,t} - \ell \frac{\partial p}{\partial y}\Big|_{y=0,t} &= -v.
\end{align}
\end{linenomath}
We discretise space into $N+1$ nodes at positions $y_0,\ldots,y_N$ uniformly spaced with spacing $\Delta y$, and use centered finite differences to evaluate spatial derivatives:
\begin{linenomath}
\begin{equation}\label{eq:FD}
  \frac{dp_n}{dt} = \frac{p_{n+1} - 2p_n + p_{n-1}}{\Delta y^2},
\end{equation}
\end{linenomath}
valid for $n=0,\ldots,N-1$, and boundary conditions read
\begin{linenomath}
\begin{equation}\label{eq:BCtop_FD}
  \frac{dp_N}{dt} = 0
\end{equation}
\end{linenomath}
and
\begin{linenomath}
\begin{equation}\label{eq:BCbottom_FD_prep}
  \frac{dp_0}{dt} - \ell \frac{p_1 - p_{-1}}{2\Delta y} = h'[\delta]v/\delta_0,
\end{equation}
\end{linenomath}
where a ghost node at $n=-1$ was introduced. Combining \eqref{eq:BCbottom_FD_prep} into \eqref{eq:FD} at $n=0$, we find the time derivative of pore pressure in the fault as
\begin{linenomath}
\begin{equation}\label{eq:BCbottom_FD}
  \frac{dp_0}{dt} = \left[\frac{p_1-p_0}{\Delta y}+\frac{h'[\delta]v/\delta_0}{\ell}\right]/(\Delta y/2 + 1/\ell).
\end{equation}
\end{linenomath}

Equations \eqref{eq:ode1}, \eqref{eq:ode2}, \eqref{eq:FD} for $n=1,\ldots,N-1$, \eqref{eq:BCtop_FD} and \eqref{eq:BCbottom_FD} form a system of $N+3$ ODEs in time for unknowns $\{v,\delta,p_0,\ldots,p_N\}$. We solve this system using a 5th order, A-L stable, stiffly-accurate, explicit singly diagonal implicit Runge-Kutta method with splitting (see \citet{Kennedy2003}, method implemented as \verb+KenCarp5()+ in the \verb+DifferentialEquations.jl+ package, see \citet{Rackauckas2017}).

All physical parameters, with the exception of the storage $S_\mathrm{f}w$ and characteristic cohesion-weakening distance $\delta_\mathrm{c}$, have been measured or imposed during the shear failure experiments: Intact hydraulic parameters $\kappa$ and $S$ were measured prior to deformation (Figure S1), the shear stress drop and coefficient of friction were obtained from mechanical data, and $P_\mathrm{c}$ and $p_0$ were imposed (Table \ref{tab:1}). 

We assume that dilation is linear with slip, and we set the dilation rate to zero for fault slip beyond the cohesion weakening distance $\delta_\mathrm{c}$, thereby assuming that dilation is largest during the formation of the fault and neglecting dilation during fluid recharge driven slip. This is supported by the order of magnitude difference between dilation rates measured for failure and for slip (Figure \ref{fig:porevolumechange}). To ensure that the total increase in pore volume after shear failure in the simulations is equal to the measured increase in pore volume, we normalised the dilation rate $dw$ (i.e., increase in fault zone width) by the input value used for $\delta_\mathrm{c}$. Hence, the dilation rates for shear failure given in Table \ref{tab:1} are a lower bound. Note that more elaborate functions to describe dilation with slip have been proposed \citep[e.g.,][]{Rudnicki1988, Segall1995}, based on dilation rates measured in gouge. However, for shear failure of intact rock, dilation was only measured afterwards, so that a linear increase during failure and subsequent slip is the most straigthforward approximation. 

We use the spring-slider model to simulate the slip rate and pore pressure during shear failure for a range of values for $\delta_\mathrm{c}$ and $S_\mathrm{f}w$. We then calculate the misfit between the observed maximum slip rate and minimum pore pressure during shear failure (Table \ref{tab:1}) and the simulated maximum slip rate and minimum pore pressure, using a least absolute criterion and assuming a Laplacian probability density function \citep{Tarantola2005}. We use uncertainties of 0.1~mm~s$^{-1}$ and 5~MPa for the slip rate and pore pressure, respectively. From the resulting probability density, we pick the best fitting pair of values for $S_\mathrm{f}w$ and $\delta_\mathrm{c}$ (Figure \ref{fig:model}d, Table \ref{tab:1}). For the simulation results of the other stable ruptures, see Figure S4.

\paragraph{Pore pressure recovery versus slip for triaxial experiments:}
For the relation between pore pressure recovery and slip after the largest drop in pore pressure (Figure \ref{fig:recharge}) we consider the fault strength is given as $\tau = \mu(\sigma_\mathrm{n}-p)$, and is in balance with the imposed load $-k\delta$. Using the triaxial relations for shear and normal stress and machine stiffness (equations \eqref{eq:triax} and \eqref{eq:machinestiffness}), we obtain:
\begin{linenomath}
\begin{equation}
\label{eq:pslip}
p = \delta k \left( \frac{1}{\mu} - \frac{(1 - \cos(2\psi))}{\sin(2\psi)} \right). 
\end{equation}
\end{linenomath}
With $E \approx 40$~GPa, measured during unloading of the sample at the end of the experiment, $L = 5$~cm, $\psi = 30^{\circ}$, $k' = 382$ GPa~m$^{-1}$, and $\mu = 0.6$, we observe that this relation follows the slip versus pore pressure data from the stable failure experiments, where deviations for samples WG4 and WG5 are explained by variations in $\mu$, which are near to 0.7 (Figure \ref{fig:recharge}).

\clearpage
\renewcommand\thefigure{S\arabic{figure}}
\setcounter{figure}{0}

\begin{figure}
\centering
\includegraphics{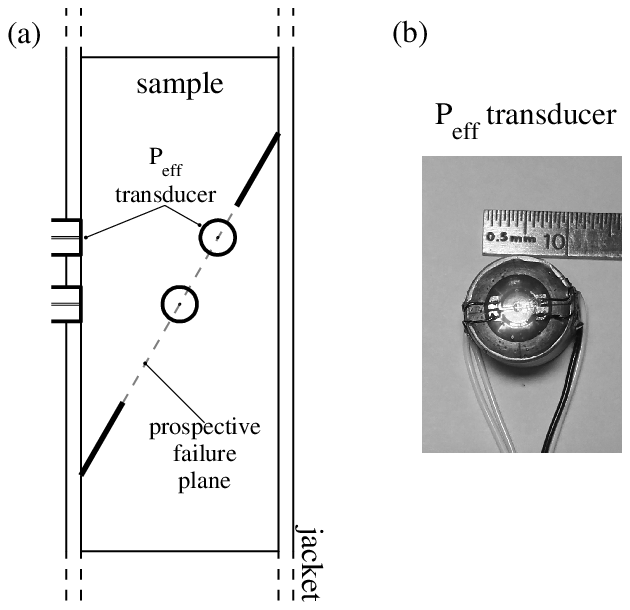}
\caption{\textbf{(a)}: Schematic of the sample setup, with two effective pressure transducers on the prospective failure plane, and two transducers on the hanging block of the sample. \textbf{(b)}: Photo of an effective pressure transducer. }
\label{fig:A}
\end{figure}

\begin{figure}
\centering
\includegraphics{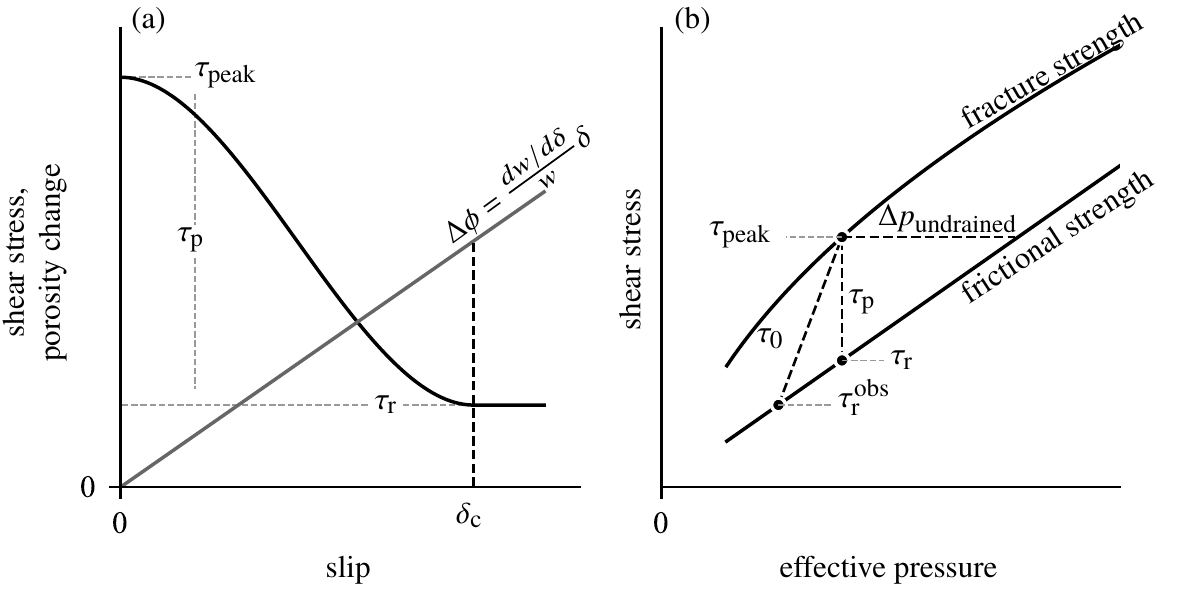}
\caption{\textbf{(a)}: Sketch of the cohesion shear stress drop $\tau_\mathrm{p}$ from peak stress $\tau_\mathrm{peak}$ down to residual frictional strength $\tau_\mathrm{r}$ (black curve) for a constant normal stress and a linear increase in porosity (gray curve) as functions of slip. \textbf{(b)}: The components of the resistance to sliding on the fault depicted in a normal stress versus shear stress sketch. One component is the cohesion shear stress drop $\tau_\mathrm{p}$ for a constant normal stress. The normal stress is not constant for triaxial deformation experiments, so that the observed shear stress drop $\tau_\mathrm{0}$ is larger than $\tau_\mathrm{p}$. The other component is caused by a localised pore pressure reduction, here we show an example of how the undrained isothermal pore pressure drop $\Delta p_\mathrm{undrained}$ increases the resistance to frictional sliding on the fault. }
\label{fig:ss_sketch}
\end{figure}

\begin{figure}
\centering
\includegraphics{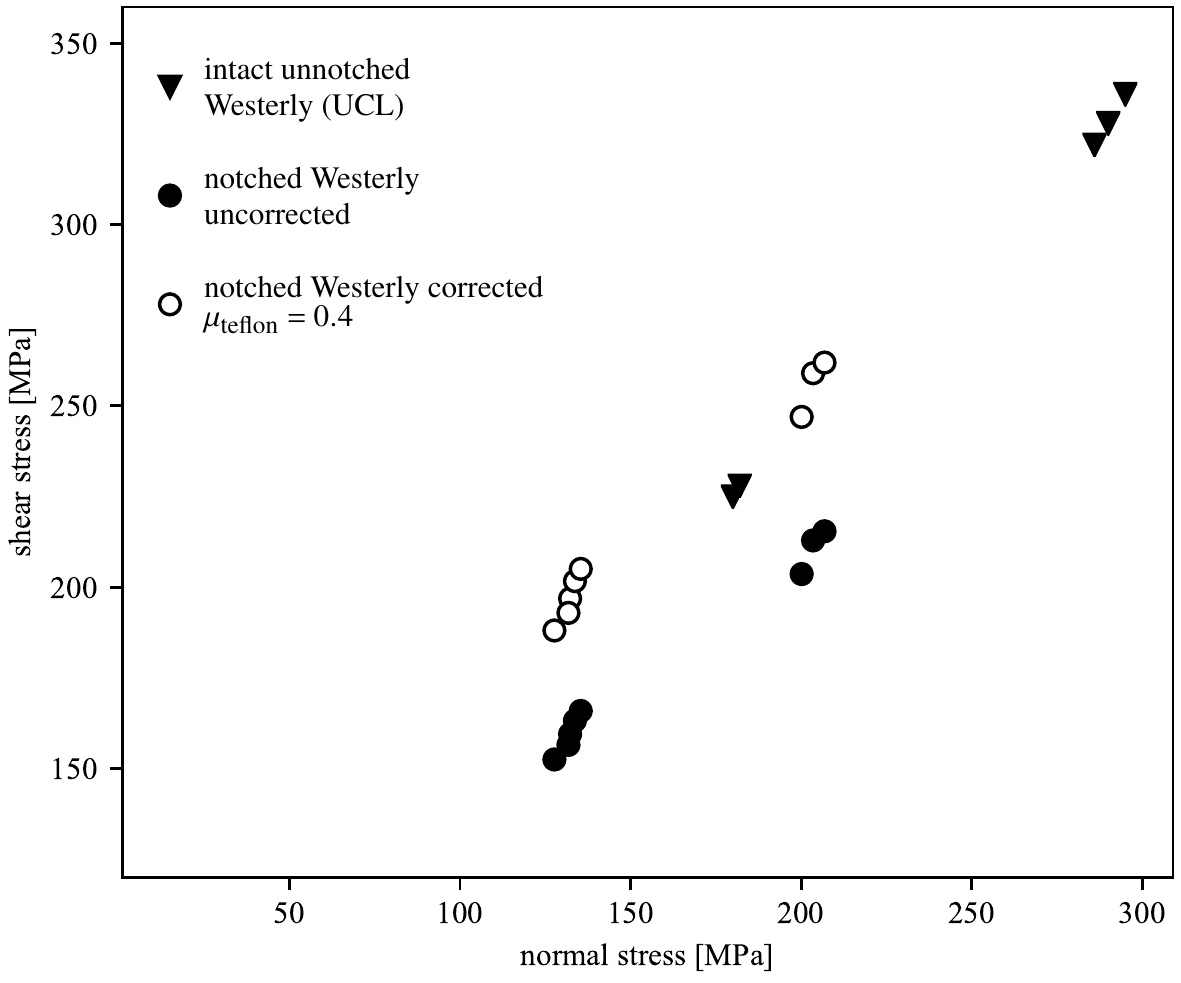}
\caption{Peak stresses measured on intact Westerly granite with notches (round filled marks, this study) and on intact Westerly granite samples without notches (triangular markers, unpublished data). The unnotched samples were tested at 50 and 100 MPa confining pressure. Both sets of experiments where performed in the same triaxial loading apparatus, and samples were cored from the same granite blocks. The peak shear stresses of the notched samples are matched with those of the intact samples for a shear stress correction with a rock-Teflon-rock friction coefficient of 0.4. }
\label{fig:shearcorrection}
\end{figure}

%\section{Figure S3: Hydraulic characterisation of undeformed thermally cracked Westerly granite}
\begin{figure}
\centering
\includegraphics{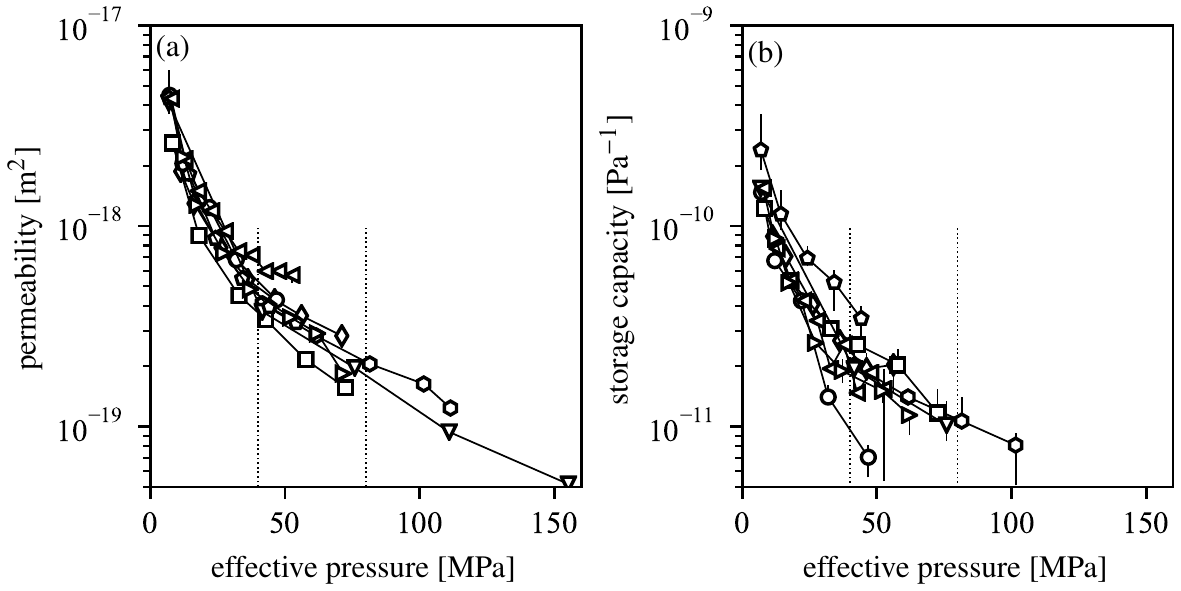}
\caption{Permeability $\kappa$  \textbf{(a)} and storage capacity $S$ \textbf{(b)} as a function of effective pressure, measured on undeformed thermally cracked Westerly granite using the transient pore pressure step method. Fitted exponential functions describing effective pressure dependence are shown as grey curves. Dashed vertical lines at experimental imposed effective pressures of 40 and 80 MPa. Left-pointing triangle = WG3, diamond = WG4, circle = WG5, square = WG7, inverted triangle = WG8, hexagon = WG10, right-pointing triangle = WG12. See Table \ref{tab:1} for sample reference.}
\label{fig:permstorativity}
\end{figure}

%\section{Figure S4: Mechanical data}
\begin{figure}
\centering
\includegraphics{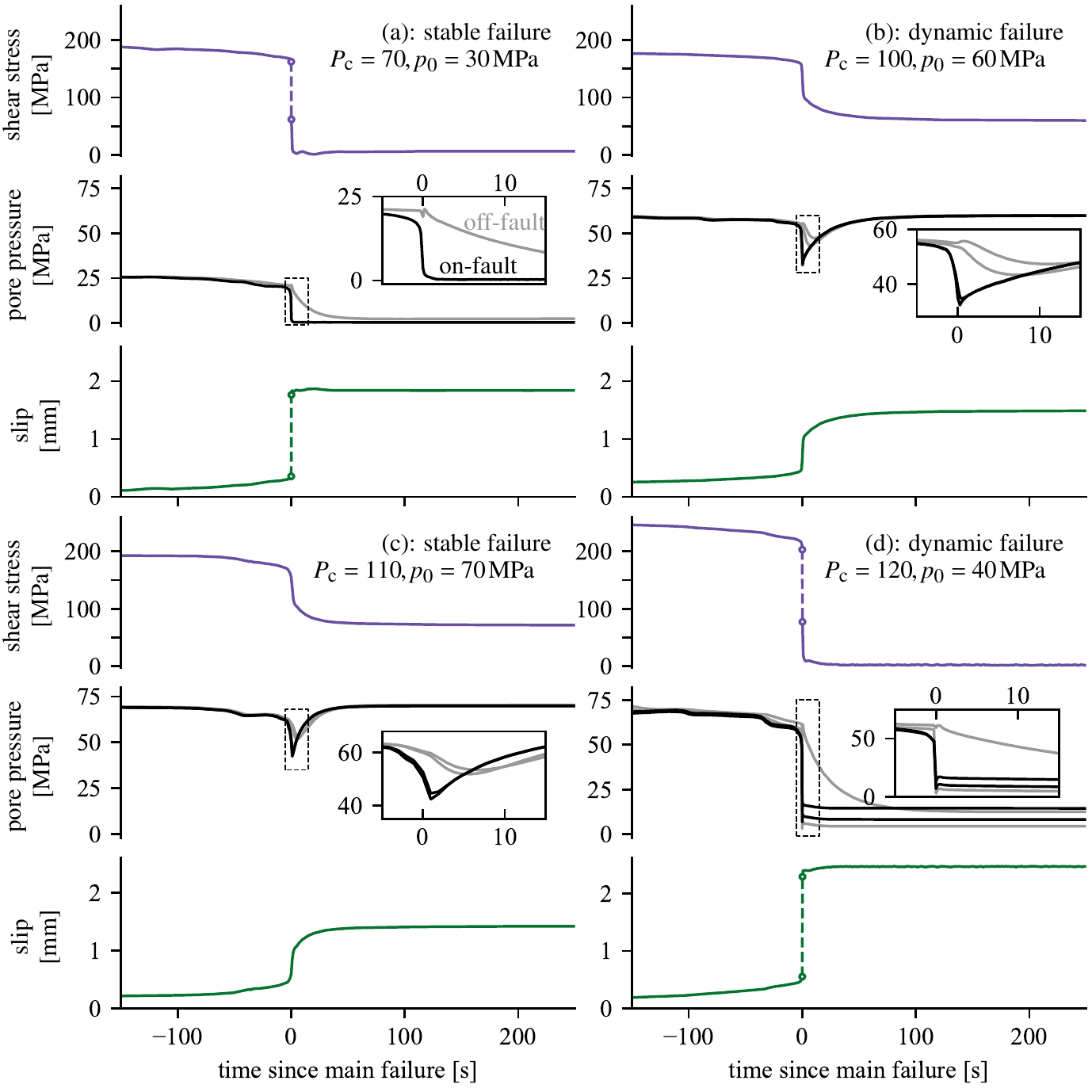}	
\caption{Pore pressure records during \textbf{(a)} dynamic failure (sample WG2, $P_\mathrm{c}=70$~MPa, $p_0=30$~MPa), \textbf{(b)} and \textbf{(c)} stable failure (sample WG5, $P_\mathrm{c}=100$~MPa, $p_0=60$~MPa, and WG12, $P_\mathrm{c}=110$~MPa, $p_0=70$~MPa), and \textbf{(d)} dynamic failure (sample WG7, $P_\mathrm{c}=160$~MPa, $p_0=80$~MPa). See Table \ref{tab:1} for sample reference.}
\label{fig:mechrecord_appendix}
\end{figure}

%\section{Figure S5: Measurements of dilation rate with slip}
\begin{figure}
\centering
\includegraphics{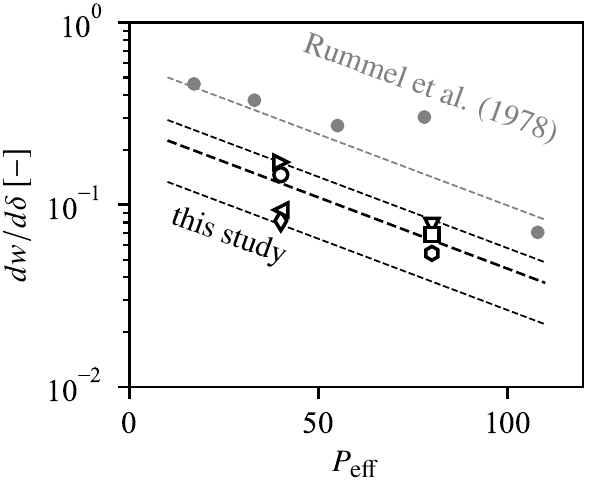}	
\caption{Dilation rate with slip for intact samples versus effective pressure measured in this study (black symbols) and extracted from shear failure data of intact Fichtelgebirge granite from \citet{Rummel1978}. The effective pressure dependency of $dw/d \delta$ is estimated by an exponential relation $dw/d \delta  = A \times \exp(-0.018 P_\mathrm{eff})$, where the constant $A = 0.27$ for thermally cracked Westerly granite (black curve) and $A = 0.6$ for Fichtelgebirge granite (gray curve). Upper and lower bound estimates for Westerly granite (thin black curves) are $A = 0.35$ and $A=0.16$, respectively. Left-pointing triangle = WG3, diamond = WG4, circle = WG5, square = WG7, inverted triangle = WG8, hexagon = WG10, right-pointing triangle = WG12.}
\label{fig:Rummel_fit}
\end{figure}

%\section{Figure S7: Spring slider model simulation results}
\begin{figure}
\centering
\includegraphics{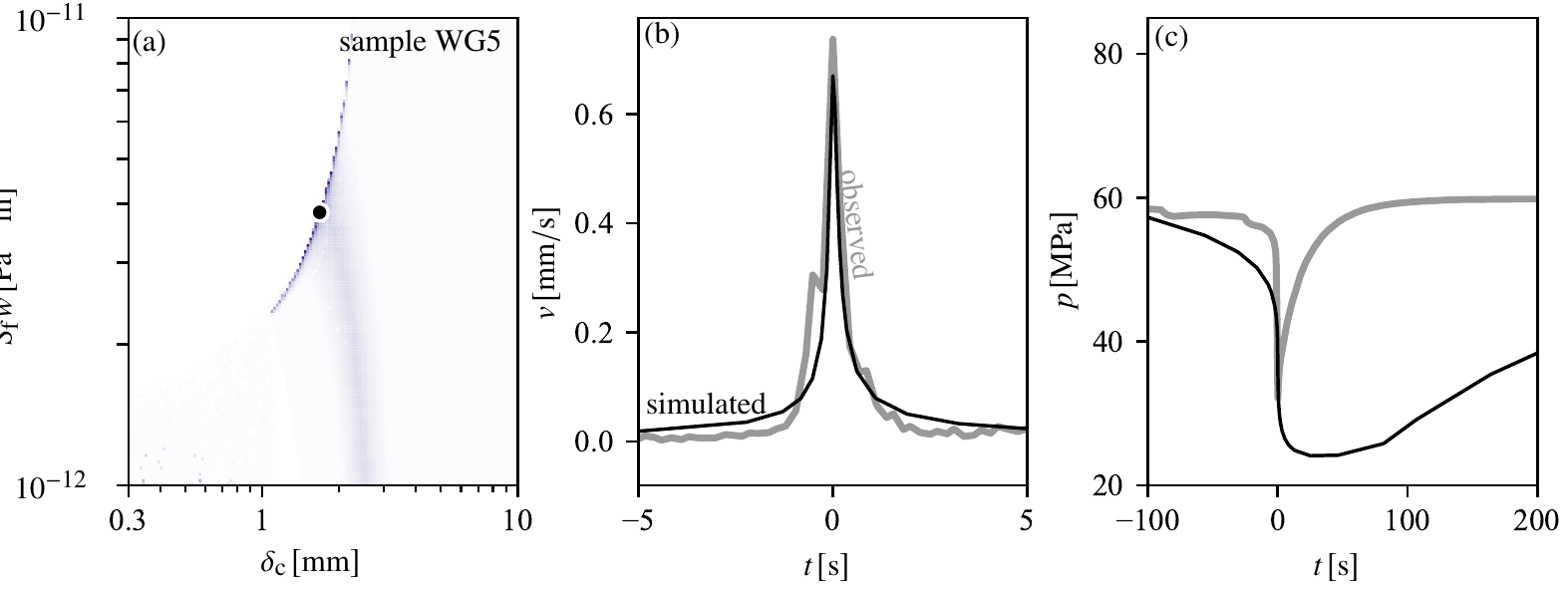}
\includegraphics{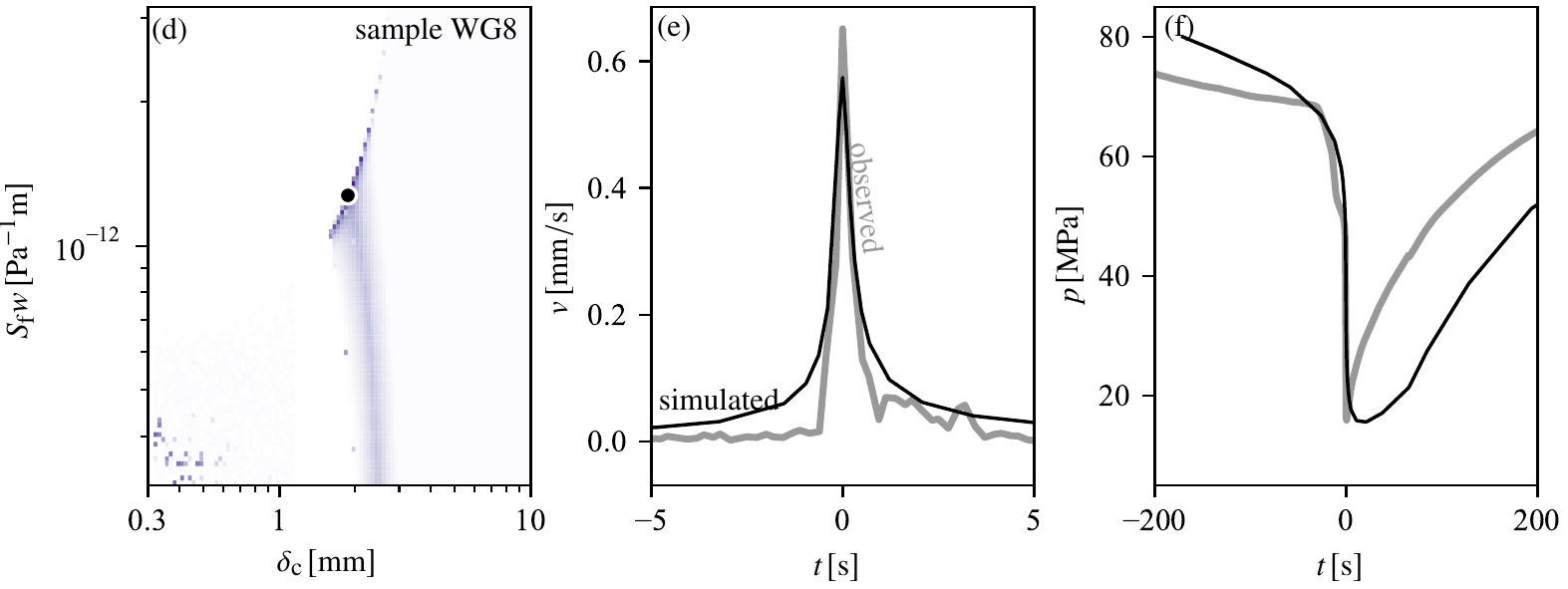}
\includegraphics{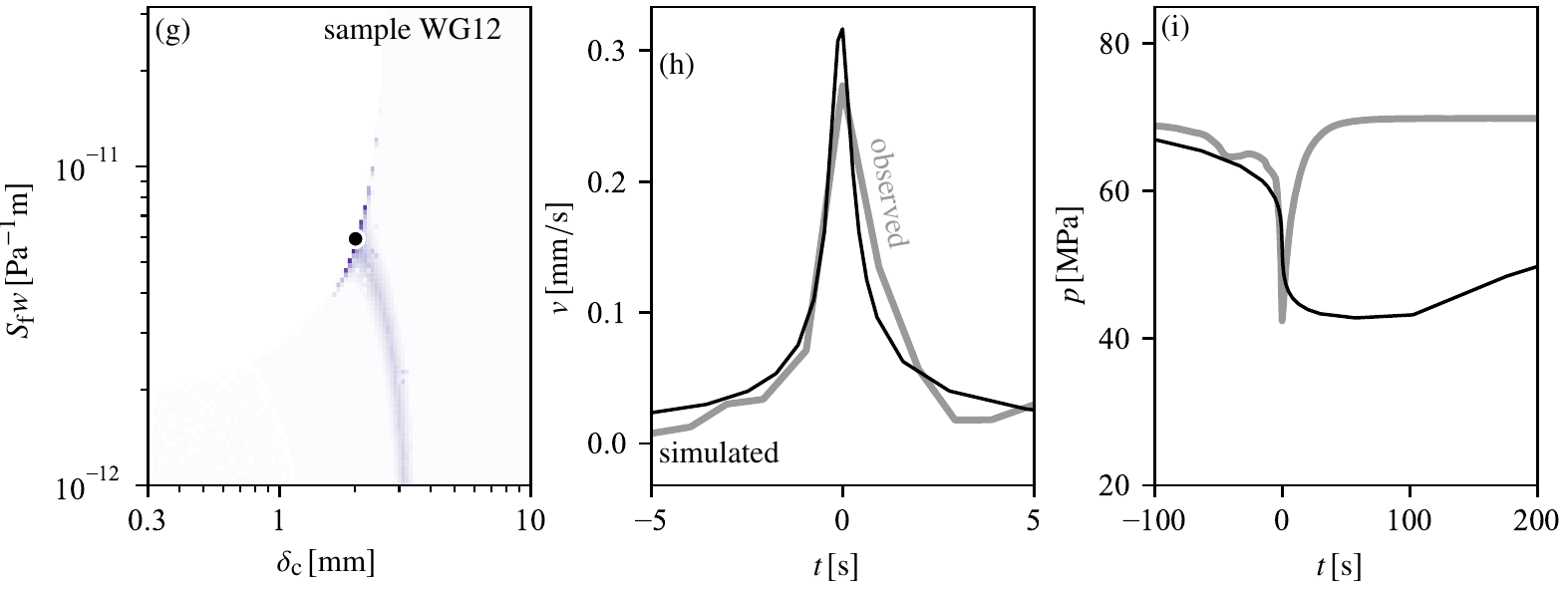}
\caption{Spring-slider simulation results for stable shear failure experiment WG5 (top), WG8 (centre), and WG12 (bottom). \textbf{(a)}: Probability density resulting from exploring $(\delta_\mathrm{c}, S_\mathrm{f}w)$-space, computed using a least absolute value criterion for the misfit between observed and simulated data. Best fitting simulation indicated by black marker. \textbf{(b)}: Observed (gray curve) and simulated (black) slip rate over time. \textbf{(c)}: Observed (gray curve) and simulated (black) fault zone pore pressure. Inset: Magnified pore pressure record around the stable failure event.}
\label{fig:dc_wSf}
\end{figure}

%\section{Figure S8: Undrained pore pressure drop and critical stiffness}
\begin{figure}
\centering
\includegraphics{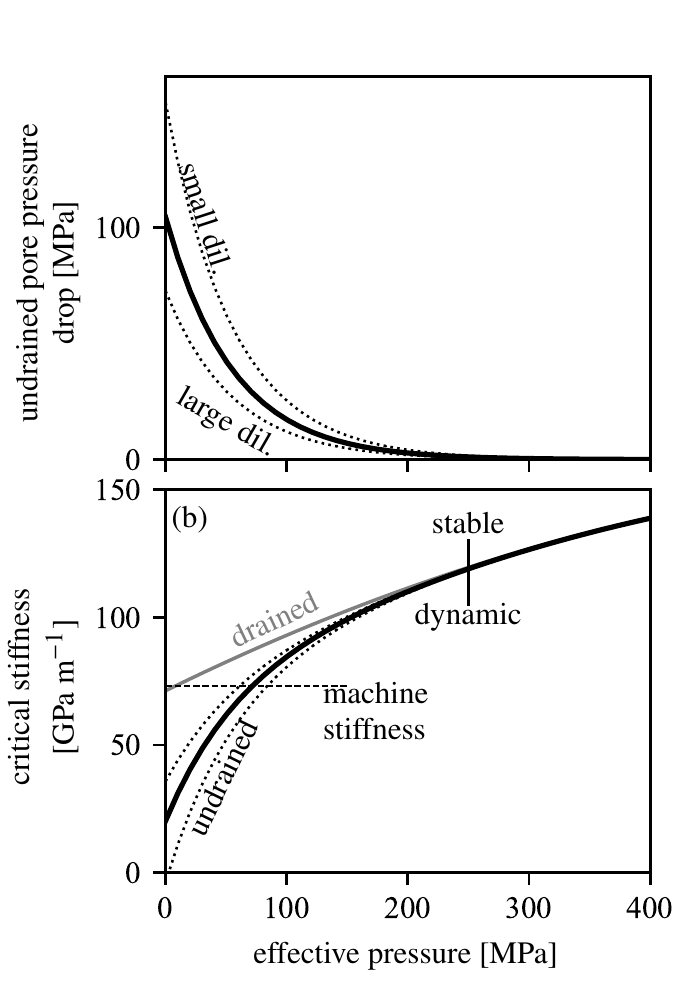} 
\caption{\textbf{(a)} Undrained isothermal pore pressure drop $\Delta p_\mathrm{undrained}$ and \textbf{(b)} critical stiffness $k_\mathrm{cr}$ versus effective pressure. Solid black line is based on the average of the estimates for $S_\mathrm{f}w$, the large and small dilation curves (dashed) are based on minimum and maximum estimates for $S_\mathrm{f}w$ and $d w/d \delta$ (Table \ref{tab:1}). Drained critical stiffness in gray. Machine stiffness is given by equation \ref{eq:machinestiffness}.}
\label{fig:pstar}
\end{figure}

\end{document}